\documentclass[10]{article}

\usepackage{amssymb,amsmath,amsfonts}
\usepackage{xspace}
\usepackage[T1]{fontenc}

\usepackage{array}
\usepackage{graphics}
\usepackage{multirow}
\usepackage{graphicx}
\usepackage[
  breaklinks=true,
  unicode=true,
  urlcolor = blue,
  colorlinks = true,
  citecolor = red,
  linkcolor = blue
]{hyperref}
\usepackage{breakurl}
\usepackage{tabularx}
\usepackage{epsfig}
\usepackage{epstopdf}
\usepackage{xspace}
\usepackage{verbatim}
\usepackage[font=small,labelfont=bf]{caption}
\usepackage{color}
\usepackage[left=1.25in,right=1.25in,top=1in,bottom=1in]{geometry}

\usepackage{nicefrac}
\usepackage{multicol}
\usepackage{placeins}
\usepackage{morefloats}

\usepackage{subcaption}
\usepackage[utf8x]{inputenc}
\usepackage[T1]{fontenc}

\newcommand{\ignore}[1]{}
\newcommand{\revised}[1]{}

\newcommand{\TODO}[1]{{{\textbf{\bf \underline{Todo:}} {\footnotesize \color{red}#1}}}}

\renewcommand{\paragraph}[1]{\medskip\noindent{\bf #1}~}

\newcommand{\namedref}[2]{\hyperref[#2]{#1~\ref*{#2}}}

\usepackage{authblk}
\usepackage{natbib}
\setcitestyle{square}

\let\originalparagraph\paragraph
\renewcommand{\paragraph}[1]{\originalparagraph{#1.}}

\usepackage[hang,flushmargin]{footmisc}

\title{\textbf{\Huge Privacy in the Genomic Era}}
\date{}
\author{Muhammad Naveed\thanks{University of Illinois at Urbana-Champaign. Work done in part at Ecole Polytechnique Federale de Lausanne. Email: naveed2@illinois.edu},
Erman Ayday\thanks{Bilkent University. Work done at Ecole Polytechnique Federale de Lausanne. Email: erman@cs.bilkent.edu.tr},
Ellen W. Clayton\thanks{Vanderbilt University. Email: ellen.clayton@vanderbilt.edu},
Jacques Fellay\thanks{Ecole Polytechnique Federale de Lausanne. Email: jacques.fellay@epfl.ch},

Carl A. Gunter\thanks{University of Illinois at Urbana-Champaign. Email: cgunter@illinois.edu},
Jean-Pierre Hubaux\thanks{Ecole Polytechnique Federale de Lausanne. Email: jean-pierre.hubaux@epfl.ch},
Bradley A. Malin\thanks{Vanderbilt University. Email: b.malin@vanderbilt.edu},
XiaoFeng Wang\thanks{Indiana University at Bloomington. Email: xw7@indiana.edu}}
\begin{document}

\makeatletter
\renewcommand*{\@fnsymbol}[1]{\ifcase#1\or*\else\@arabic{#1}\fi}
\makeatother

\maketitle

\let\cite\citep
\begin{abstract}
Genome sequencing technology has advanced at a rapid pace and it is now
possible to generate highly-detailed genotypes inexpensively. The collection
and analysis of such data has the potential to support various applications,
including personalized medical services. While the benefits of the genomics
revolution are trumpeted by the biomedical community, the increased
availability of such data has major implications for personal privacy; notably
because the genome has certain essential features, which include (but are not
limited to) \textit{(i)} an association with traits and certain diseases,
\textit{(ii)} identification capability (e.g., forensics), and \textit{(iii)}
revelation of family relationships. Moreover, direct-to-consumer DNA testing
increases the likelihood that genome data will be made available in less
regulated environments, such as the Internet and for-profit companies. The
problem of genome data privacy thus resides at the crossroads of computer
science, medicine, and public policy. While the computer scientists have
addressed data privacy for various data types, there has been less attention
dedicated to genomic data. Thus, the goal of this paper is to provide a
systematization of knowledge for the computer science community. In doing so,
we address some of the (sometimes erroneous) beliefs of this field and we
report on a survey we conducted about genome data privacy with biomedical
specialists. Then, after characterizing the genome privacy problem, we review
the state-of-the-art regarding privacy attacks on genomic data and strategies
for mitigating such attacks, as well as contextualizing these attacks from the
perspective of medicine and public policy. This paper concludes with an
enumeration of the challenges for genome data privacy and presents a framework
to systematize the analysis of threats and the design of countermeasures as the
field moves forward. 
\end{abstract}

\section{Introduction}
\label{sec:intro}

The genomic era began with the announcement twelve years ago that the Human Genome Project (HGP) had completed its goals~\cite{GuttmacherC03}. The technology associated with genome sequencing has progressed at a rapid pace, and this has coincided with the rise of cheap computing and communication technologies. Consequentially, it is now possible to collect, store, process, and share genomic data in a manner that was unthinkable at the advent of the HGP. In parallel with this trend there has been significant progress on understanding and using genomic data that fuels a rising hunger to broaden the number of individuals who make use of their genomes and to support research to expand the ways in which genomes can be used. This rise in the availability and use of genomic data has led to many concerns about its security and privacy. These concerns have been addressed with efforts to provide technical protections and a corresponding series of demonstrations of vulnerabilities. Given that much more research is needed and expected in this area, this seems like a good point to overview and systematize what has been done in the last decade and provide ideas on a framework to aid future efforts.

%The genome corresponds to the entirety of a person's DNA sequence.  The notion of DNA is not new;
%%seminal
%breakthroughs in biochemistry in the 1890s (thanks to Miescher) and biophysics in the 1950's (thanks to Crick, Franklin, Watson, and Wilkins) demonstrated its molecular makeup and structure (i.e., the famed ``double helix'').
%%, discoveries which led to the Nobel Prize.
%DNA sequencing is not a novel concept either; Sanger introduced a methodology for doing so
%%sequencing the genome
%in the 1970s.  So, why
%%is it now that we
%%are we now
%write a Systematization of Knowledge (SoK) on the topic of genome data privacy?  In short, it is because the technology associated with genome sequencing has progressed at a rapid pace -- a progression which has coincided with the rise of cheap computing technologies. Consequentially, it is now possible to collect, store, process, and share genomic data in a manner that was unthinkable less than only a decade ago.\vspace{-5pt}
%%Genome is the entirety of an organism's hereditary information.
%%It is encoded in DNA molecule, a small molecule that is discovered by Watson and Crick in
%%1953.

To provide context, consider that it was not until the early 1990s when sequencing the human genome was posited as a scientific endeavor.
The first attempt for \emph{whole genome sequencing}\footnote{In this study,
	we refer to the process of obtaining the Whole Genome Sequence (WGS)
or the Whole Exome Sequence (WES) as \emph{sequencing} and the process of
obtaining the variants (usually only single nucleotide polymorphisms, or SNPs) as \emph{genotyping}.} (a laboratory process that maps the full DNA sequence of an individual's genome) was initiated at the U.S. National Institutes of Health (NIH) in 1990 and the first full sequence was released 13 years later
at a total cost of \$3 billion. Yet, sequencing technology has evolved
%and important breakthroughs in genomic research,
and costs have plummeted, such that
%over the past 10 years, and today the
the price for a whole genome sequence is
%only
\$5K\footnote{\url{http://www.genome.gov/sequencingcosts/}} as of July 2014 and can be completed in two to three days.
%It is expected that
%Cheaper and faster technology will soon make
The ``\$1K genome in 1 day'' will soon be a reality.

%Moreover,
Decreases in sequencing costs have coincided with an escalation in genomics as a research discipline with explicit application possibilities. Genomic data is increasingly incorporated in a variety of domains, including healthcare (e.g., personalized medicine), biomedical research (e.g., discovery of novel genome-phenome associations), %(e.g., genome-wide association studies),
direct-to-consumer (DTC) services (e.g., disease risk tests),
% or ancestry search),
and forensics (e.g., criminal investigations).
%legislation (e.g., paternity test),
For example, it is now possible for physicians to prescribe the ``right drug at the right time'' (for certain drugs) according to the makeup of their patients' genome~\cite{bielinski2014preemptive,overby2010feasibility,gottesman2013clipmerge,pulley2012operational}.
%Furthermore, individuals can learn their genetic predispositions to serious diseases, or couples can find out if their potential offspring will carry specific genetic diseases.

To some people, genomic data is considered (and treated) no differently than traditional health data (such as what might be recorded in one's medical record)
%electronic health records (EHRs)
or any other type of data more generally~\cite{bains2010genetic,rothstein2005genetic}.
%``big data''.
While genomic data may not be ``exceptional'' in its own right, it has many features that distinguish it
%from other types of data
(discussed in depth in the following section) and there is a common belief that it should be handled (e.g., stored, processed, and managed) with care.
%It may be that no
%In fact, we believe that today, no other type of data lies on the intersection of such many unique features that genomic data has.
%It is true that due to some of these features, genomic data are prominent, especially for healthcare (as it has been paving the way to personalized medicine).
The privacy issues associated with genomic data are complex, particularly because such data has a wide range of uses and provides information on more than just the individual from which the data was derived.
%For instance, it is possible that the leakage of genomic data may have consequences for discrimination (e.g., for health insurance) or blackmail.
Yet, perhaps most importantly, there is a great fear of the unknown.  Every day, we learn something new about the genome, whether it be knowledge of a new association with a particular disease or proof against a previously reported association.  We have yet to discover everything there is from DNA, which makes it almost impossible to assign exact value, and thus manage DNA as a personal asset (or public good).
%However, again, due to some of these
%features, genomic data causes a severe privacy risk for
%individuals (unless it is handled with care).
%For example, genomic data includes information about an individual's predisposition to several privacy-sensitive diseases (e.g., Alzheimer's), and more importantly, genomic data of an individual also contains information about his family members.
%Furthermore, due to the immaturity of genomic research, even the geneticists do not know everything about the genome. For instance, it is yet unclear to what extent genomic information will be able to predict diseases or drug responses.
%As a consequence,
So, as the field of genomics evolves, so too will the views on the privacy-sensitivity of genomic data.
%change.
%keep increasing.
%Whereas, we observe that the importance of genome privacy, including the aforementioned consequences and the extend of these threats are still not fully understood by many people.
As this paper progresses, we
%will
review some of the common beliefs revolving around genome privacy. And, in doing so, we
%will
report on the results of a survey we conducted with biomedical specialists regarding their perspective on genome data privacy issues.

It should be recognized that there exist numerous publications on technical, ethical, and legal aspects of genomics and privacy. The research in the field covers privacy-preserving handling of genomic data in various environments (as will be reviewed in this paper).
%
%from several disciplines in order to protect the privacy of genomic data (that we summarize in detail in this paper).
%Even though existing work brings a lot of solutions to today's problems,
Yet, there are several challenges to ensuring that genomics and privacy walk hand-in-hand.  One of the challenges that computer scientists face is that these views tend to be focused on one aspect of the problem in a certain setting with a certain discipline's perspective. From the perspective of computer science, there is a need for a complete framework which shows \textit{(i)} what type of security and privacy requirements are needed in each step of the handling of genomic data, \textit{(ii)} a characterization of the various threat models that are realized at each step, and \textit{(iii)} open computational research problems. By providing such a framework in this paper,
%also create such a framework to
we are able to illustrate the important problems of genome privacy to computer science researchers working on security and privacy problems more generally.
%Most importantly,

\paragraph{\textbf{Related Surveys and Articles}}
Privacy issues caused by forensic, medical, and other uses of genomic data have
been studied in the past few
years~\cite{Stajano2008,Stajano2009,malin2005evaluation,survey2013,Naveed2014,de2014genomic}.
A recent survey~\cite{erlich2013} discusses privacy breaches using genomic
data and proposes methods for protection. It addresses topics that we discuss
in Sections~\ref{sec:riskassess} and Section~\ref{sec:framework} of this paper.
In Section~\ref{sec:framework} we present an end-to-end picture for the
handling of genomic data in a variety of contexts as shown in
Figure~\ref{fig:flow}, while \cite{erlich2013} discusses how access control,
data anonymization and cryptographic techniques can be used to prevent genetic
privacy breaches. Moreover, \cite{erlich2013} has been written for a general
audience, whereas this paper is meant for computer scientists (and in
particular security and privacy specialists).

\paragraph{\textbf{Contributions}}
Following are the main contributions of this paper:
%summarized as follows:
\begin{itemize}
	\item We provide an \textit{extensive and up-to-date (as of June 2015) literature survey}\footnote{In this paper, the word ``survey'' is used to mean
	\emph{literature survey} as well as \emph{opinion poll}, however, the meaning
	will be clear from the context.} of computer science as well as medical
	literature about genome privacy.
%The literature survey is up to date as of June 2015.
	\item We report concerns expressed by an opportunistically ascertained group of biomedical specialists about the security and privacy of genomic data.
%	\item We report on the results of an \textit{expert survey (opinion poll)}
%%and provide its results to inform security and privacy
%%we conducted with
%%	community about the
%biomedical specialists' regarding their point of view on the security and privacy of genomic data.
	\item  We
develop an end-to-end \textit{framework for the security and privacy of genomic data} in a
variety of healthcare, biomedical research, legal and forensics, and direct-to-consumer
contexts.
%We present an almost complete picture of the security and privacy issues for genomics.
	\item We present what we believe to be the first
	document that reflects the \emph{opinions of computer
	science, medical, and legal researchers} for this important topic.
\end{itemize}

We also provide an online tutorial\footnote{Available at
\url{https://sites.google.com/site/genoterms}} of biology and other related material
%tailored for computer scientists to
to define technical terms used in this (and other) paper(s)
%or other on papers
on the security and privacy of genomic data. The remainder of this paper is organized as follows. Section~\ref{sec:whyexceptional} explains to what extent genomic data is distinct
%(though not completely different)
from data in general and health information in particular.
%and other forms of ``big data''.
Section~\ref{sec:background} provides an overview of uses of genomic
data for the non-specialist. Section~\ref{sec:importance} emphasizes the
relevance of genome privacy. Section~\ref{sec:survey} reports on the
concerns of 61 opportunistically ascertained biomedical scientists regarding the importance of genomic data privacy and security.
% in various settings.
%\COMMENT{Survey is not only for beliefs}
%associated with genomics and privacy are held by practitioners, we conducted asurvey with ~61 biomedical specialists, located primarily in the US and in Europe.
%this survey.
%A certain mystique has developed around genome privacy and many people have developed several beliefs. In the same section, we also analyze and assess the most frequent of these beliefs.
Sections~\ref{sec:riskassess}
and~\ref{sec:solutions} provide literature surveys, where the former
summarizes the problem (i.e., the privacy risk) and the latter summarizes
possible solutions.
%(based on various computational mechanisms, such as cryptography).
Section~\ref{sec:challenges} summarizes the challenges for
genomic medicine and privacy. Based on this analysis,
Section~\ref{sec:framework} offers a general framework for privacy-preserving
handling of genomic data,
%. This section also
including an extensive threat
model that discusses what type of attacks are possible
% (based on known publications, as well as new threat models)
%proposed attacks)
at each step of the data
flow. 
%Finally, we conclude the paper in Section~\ref{secc:conclusion}.

%\TODO{For computer reading, any item requiring explanation can be clicked to see explanation on the website. For paper reading, a mark in the margin appears for terms described on the website}
\section{Special features of genomic data}
\label{sec:whyexceptional}
%In this section, we discuss why genomic data is special. We have identified six special features of genomic data shown in Figure~\ref{fig:whyexceptional}. While many other data can have some of these special features, we believe that no other data (including other molecular data such as proteomics data) have \emph{all} of these features.
%To see this special nature of genomic data, one can try to replace \emph{DNA} in Figure~\ref{fig:whyexceptional} with other types of data and see if they have all of these features.

\begin{figure}
\centering
\includegraphics[width=0.5\textwidth]{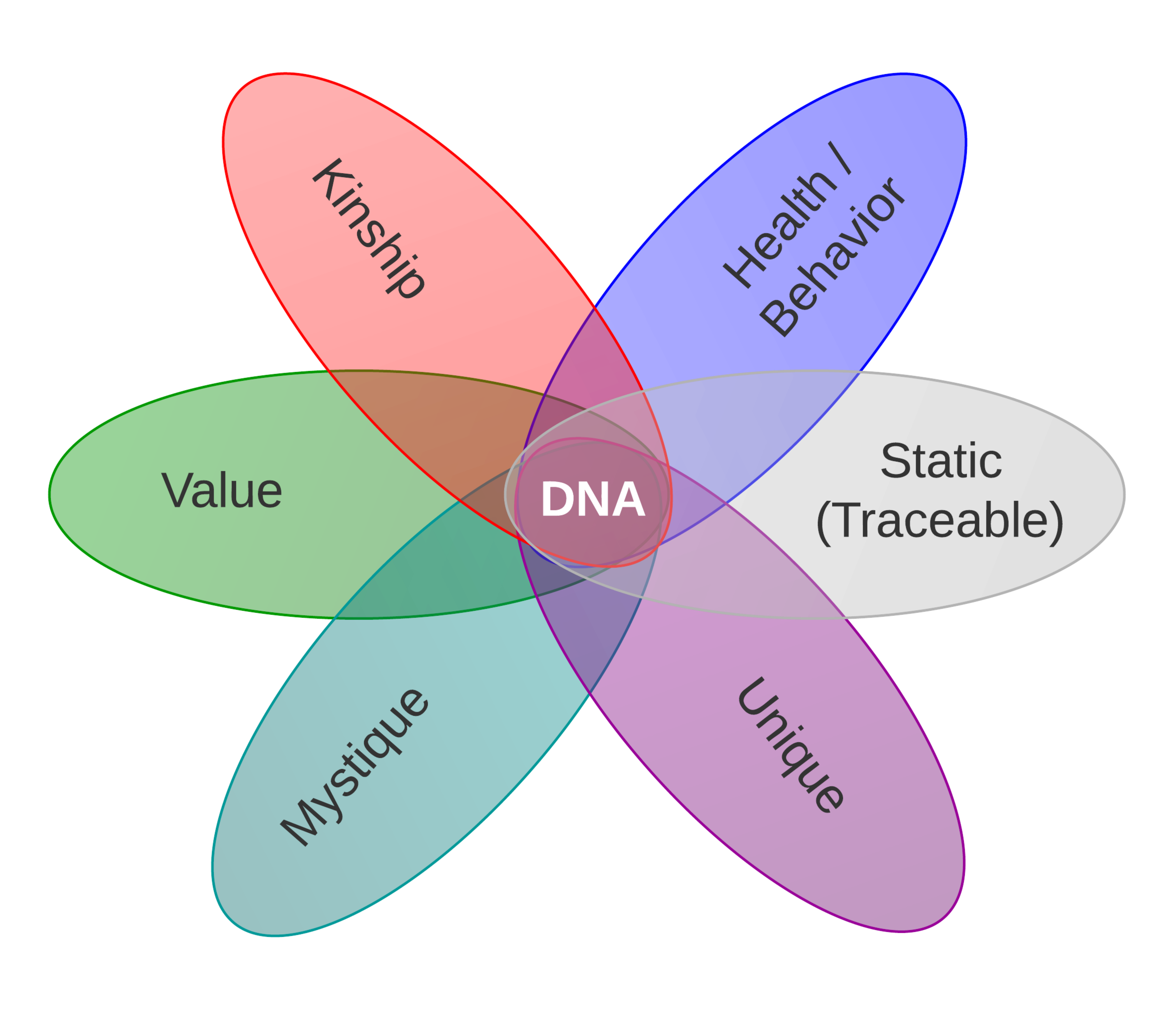}
%\caption{\footnotesize Properties of DNA that, in combination, may distinguish it from other data types.\\ \textit{Health/Behavior} means that DNA contains information about an individual health and behavior. \textit{Static(Traceable)} means that DNA does not change much over time in an individual. \textit{Unique} means that DNA data of any two individuals can be easily distinguished. \textit{Mystique} refers to the public perception of mystery about DNA. \textit{Value} refers to the importance of information content in DNA and that this importance does not decline with time (which is the case with other medical data e.g., blood pressure, glucose level, or a blood test). In fact, this importance will likely increase with time. \textit{Kinship} means that DNA contains information about an individual blood relatives.}
\caption{Properties of DNA that, in combination, may distinguish it from other data types. \textit{Health/Behavior} means that DNA contains information about an individual's health and behavior. \textit{Static(Traceable)} means that DNA does not change much over time in an individual. \textit{Unique} means that the DNA of any two individuals can be easily distinguished from one another. \textit{Mystique} refers to the public perception of mystery about DNA. \textit{Value} refers to the importance of information content in DNA and that this importance does not decline with time (which is the case with other medical data e.g., blood pressure, glucose level, or a blood test). In fact, this importance will likely increase with time. \textit{Kinship} means that DNA contains information about an individual blood relatives.}

%\caption{Why DNA is different for any other kind of data? No other data have all of the above six properties.}
\label{fig:whyexceptional}
\end{figure}

In this section, we discuss why genomic data is special. We have identified six features of genomic data, as shown in Figure~\ref{fig:whyexceptional}, and while other data harbor some of these features, we are not aware of any data (including other molecular, such as proteomics, data) that have \emph{all} of these features. 

Consider the following scenario. Alice decides to have her genome sequenced by a service called MyGenome.com that keeps her data in a repository and gives Alice information about it over time. At first she uses information from MyGenome to explore parts of her family tree and contribute her genomic data, along with some facts about herself, to support medical research on diseases of her choosing. Many years after MyGenome performed the initial sequencing, Alice began experiencing health problems for which she visited a doctor who used her genomic data to help diagnose a likely cause and customize a treatment based on variation in her genome sequence. Alice was impressed by this experience and wondered what other conditions might be in her future. After some exploration she discovered that evidence (based on published research papers) suggested a high risk of dementia for people with her genomic profile. She worried that various parties, including MyGenome, the genealogy service, and research studies with whom she shared her data, might share this and other information in ways that she did not expect or intend and whether this might have undesired consequences for her.

Alice's story highlights several of the special features of genomic data. We depict six of them in Figure~\ref{fig:whyexceptional}, which we review for orientation of the reader.
%\begin{figure}[H]
%\centering
%\includegraphics[width=0.3\textwidth]{Figures/whyexceptional.eps}
%\caption{Properties of DNA that, in combination, may distinguish it from other data types.}
%%\caption{Why DNA is different for any other kind of data? No other data have all of the above six properties.}
%\label{fig:whyexceptional}
%\end{figure}

%%%%%%%%%%%%%%%%%%%%%%%%%%%%%%%%%%%%%%%%%%%%%%%%%%%%%%%%%%
%\begin{wrapfigure}[13]{r}{5.2cm}
%%%%%%%%%%%%%%%%%%%%%%%%%%%%%%%%%%%%%%%%%%%%%%%%%%%%%%%%%%
How does the result of a DNA-based lab test differ from that of other tests? One notable feature is how it is static and of long-lived value. Most tests, especially ones Alice could do for herself, like taking her temperature and blood pressure, are of relatively short term value, whereas genomic data changes little over a lifetime and may have value that lasts for decades.  Of course, there are some exceptions to this longevity.  For instance, sequencing techniques improve in accuracy over time, so tests may be repeated to improve reliability.  Additionally, there are some modifications in DNA that accumulate over time (e.g., shortening of the ends of DNA strands due to aging~\cite{harley1990telomeres}). Most particularly, somatic mutations occur resulting in some degree of mosaicism in every individual: the most striking examples are the deleterious modifications of the DNA observed in cancer cells in comparison to DNA derived from normal cells. However, this long-lasting value means that holding and using genomic data over extended periods of time, as Alice did, is likely.

Alice's first use of her genomic data is expected to be a key driver for application development in the future. While DNA has been used for some time in parentage tests, it can be generalized from such studies to enable broader inference of kinship relations.  Services such as Ancestry.com and 23andme.com already offer kinship services based on DNA testing. While a substantial portion of Alice's DNA is in common with that of her relatives, it is also unique to her (unless she has an identical twin). This has another set of implications about potential use of genomic data, like its ability to link to her personally, a property that makes DNA testing useful in criminal forensics.

Another of the special features of DNA relates to its ability for diagnosing problems in health and behavior. Tests are able to demonstrate increased likelihood for conditions such as macular degeneration in old age and Alzheimer's (the most common form of dementia)~\cite{goldman2011genetic}. Although these are often probabilities, they can have diagnostic value as well as privacy ramifications~\cite{seddon2011risk}. For instance, if Alice's relatives learned about her increased risk of dementia, might they (consciously or unconsciously) trust her judgement a little less? Or might they instead help her to get timely treatment? This power for good and bad has led genomic data to have a certain ``mystique'', which has been promoted by scientists and the media~\cite{tambor2002}. %The ``mystique'' surrounding the genomic data is also evident from movies and books on the topic. Examples include the movie ``GATACCA'' and the book ''The DNA mystique''~\cite{nelkin1995dna}.
The ``mystique'' surrounding the genomic data is evident from movies and books on the topic. Examples include the movie “GATTACA” and the book ”The DNA mystique”~\cite{nelkin1995dna}.

Although there are many other types of tests (e.g., protein sequence tests) that carry key common information with DNA tests, there is a special status that DNA data has come to occupy, a status that some have phrased as ``exceptional'' \cite{bains2010genetic}. These special fears about the sharing of genomic data, whether founded or not, cannot be ignored when considering privacy implications. Hence, while DNA data may or may not be exceptional \cite{evans10,gostin99}, it is special in many ways, and hence warrants particular care.

\section{Uses of Genomic Data}
\label{sec:background}
An individual's genomic sequence contains over 3 billion base pairs, which are distributed across twenty-three chromosomes. Despite its size, it is estimated that the DNA of two individuals differ by no more than 0.5\% \cite{venter}; but it is these differences that influence an individual's health status and other aspects (as discussed in Section~\ref{sec:whyexceptional}).
%
%Genomic data is derived from tissue and, as such, is associated with with a wide array of applications.
To provide further context for the importance
%and uses
of genomic data, this section reviews several of the major applications in practice and under development.

%Usage of genomic data (as of today) can be classified into following three broad categories.

\subsection{Healthcare}
First, it has been recognized that mutation in an individual's genomic sequence can influence his well being. In some cases, changes in a particular gene will have an adverse effect on a person's health immediately or at some point in the future
%(i.e., the \emph{one gene, one disease}, or OGOD, model)
~\cite{botstein}. As of 2014, there were over 1,600 of these traits
%(Mendelian diseases - named after the famed geneticist Gregor Mendel)
reported on in the literature\footnote{\url{http://www.ncbi.nlm.nih.gov/Omim/mimstats.html}}, ranging from metabolic disorders (e.g., phenylketonuria, which is caused by a mutation in the PKU gene) to neurodegenerative diseases (e.g., Huntington's disease, which is caused by a mutation in the HD gene~\cite{macdonald1993novel}) to blood disorders (e.g., Sickle cell anemia, caused by a mutation in the HBB gene~\cite{saiki1985enzymatic}).  While some of these diseases are manageable through changes in diet or pharmacological treatments, others are not and have no known intervention to assist in the improvement of an individual's health status.  Nonetheless, some individuals choose to learn their genetic status, so that they may order their affairs accordingly and contribute to medical research~\cite{Mastromauro} (as elaborated upon below).  Moreover, genetic tests can be applied in a prenatal setting to detect a variety of factors that can influence health outcomes (e.g., if a fetus is liable to have a congenital defect that could limit its lifespan, such as Tay-Sach's disease)~\cite{Lippman}.

Yet, the majority of variations in an individual's genome do not follow the monogenic model. Rather, it has been shown that variation is associated with change in the susceptibility of an individual to a certain disease or behavior \cite{botstein}. Cancer-predisposing variants in genes such as BRCA1/2 or the Lynch Syndrome are well-known examples. Such variation may also modify an individual's ability to respond to a pharmaceutical agent.  For instance, some individuals are slow (fast) metabolizers, such that they may require a different amount of a drug than is standard practice, or may gain the greatest benefit from a different drug entirely.  This variation has been leveraged to provide dosing for several medications in practice, including blood thinners after heart surgery (to prevent clotting) and hypertension management (to lessen the severity of heart disease) \cite{pulley2012operational}.  Additionally, changes in an individual's genome detected in a tumor cell can inform which medications are most appropriate to treat
%an individual's
cancer \cite{mcdermott}.

%\TODO{Diagnosis, Personalized Medicine, Preventive Medicine, Pre-pregnancy Testing for healthy pregnancy and healthy-baby}

\subsection{Research}
While the genome has been linked with a significant number of disorders and variable responses to treatments, new associations are being discovered on a weekly basis.  Technology for performing such basic research continues to undergo rapid advances \cite{brunham}. The dramatic decrease in the cost of genome sequencing has made it increasingly possible to collect, store, and computationally analyze sequenced genomic data on a fine-grained level, as well as over populations on the order of millions of people (e.g., China's Kadoorie biobank~\cite{chen2011china} and UK Biobank~\cite{pmid24553384} will each contain genomic data on 500,000 individuals by the end of 2014, while the U.S. National Cancer Institute is at the beginning of its Million Cancer Genome Project~\cite{haussler2012million}).  Yet, it should be recognized that computational analysis is separate from, and more costly than, sequencing technology itself (e.g., the \$1K \emph{analysis} of a genome is far from being developed).
%\ignore{Determining sequence of nucleotides\footnote{Building block of DNA} of human DNA was the first step to unlock the treasure of information contained in human genome. The more important and more difficult task is to map the genotype information to its corresponding phenotype. This is called Genome Wide Association Studies (GWAS). Large number of people participate in GWAS and some aggregate statistics are published in the research papers.}

%\TODO{GWAS, millions of patients are sequenced (but data is kept private, if somehow other researchers can access this data it will be very useful for humanity.)}
Moreover, technological advances in genome sequencing are coalescing with a big data revolution in the healthcare domain.  Large quantities of data derived from electronic health records (EHRs), for instance, are being made available to support research on clinical phenotypes that, until several years ago, were deemed to be too noisy and complex to model~\cite{gottesman}.  As a consequence, genome sequences have become critical components of the biomedical research process~\cite{kohane:11}.

\subsection{Direct-to-Consumer Services}
Historically, genome sequencing was a complex and expensive process that, as a result, was left to large research laboratories or diagnostic services, but in the past several years, there has been a rise in DTC (Direct-to-consumer) genome sequencing from various companies~\cite{prainsack}.
%, such as 23andme.com.
These services have made it affordable for individuals to become directly involved in the collection, processing, and even analysis of their genomic data.  The DTC movement has enabled individuals to learn about their disease susceptibility risks (as alluded to earlier), and even perform genetic compatibility tests with potential partners. Moreover, and perhaps more importantly,
%the DTC movement
DTC has made it possible for individuals to be provided with digital representations of their genome sequences, such that they can control how such information is disclosed, to whom, and when.

Of course, not all consumer products are oriented toward health applications. For example, genomic data is increasingly applied to determine and/or track kinship.  This information has been applied for instance to track an individual's ancestral heritage and determine the extent to which individuals with the same surname are related with respect to their genomic variance \cite{Jobling01}.

\subsection{Legal and Forensic}
Given the static nature of genomic sequences,
%(that is, there is minimal variation across an individual's tissue and time),
this information has often been used for investigative purposes. For instance, this information may be applied in contested parentage suits~\cite{anderlik2003assessing}. Moreover, DNA found at a crime scene (or on a victim) may be used as evidence by law enforcement to track down suspected criminals~\cite{kaye2003dna}. It is not unheard of for residents of a certain geographic region to be compelled to provide tissue samples to law enforcement to help in such investigations~\cite{greely2006family}.  Given the kinship relationships that such information communicates, DNA from an unknown suspect has been compared to relatives to determine the corresponding individual's likely identity in order to better facilitate a manhunt.

One of the concerns of such uses, however, is that it is unclear how law enforcement may retain and/or use this information in the future.  The U.S.~Supreme Court recently ruled that it is permissible for law enforcement to collect and retain DNA on suspects, even if the suspects are not subsequently prosecuted~\cite{2013maryland}. Once DNA is shed by an individual (such as from saliva left on a coffee cup in a restaurant) it has been held as an ``abandoned'' resource~\cite{Joh2006}, such that the corresponding individual relinquishes rights of ownership.  While the notion of ``abandoned DNA'' remains a hotly contested issue, it is currently the case in the U.S. that DNA collected from discarded materials can be sequenced and used by anyone without the consent of the individual from which it was derived. 
\section{Relevance of Genome Privacy}
\label{sec:importance}

As discussed in Section~\ref{sec:whyexceptional}, genomic data has numerous distinguishing features and applications.
%Therefore, privacy of genetic information is very important as it can have long lasting impact on an individual's life. One's genome contains privacy-sensitive information such as his predisposition to several diseases, his ethnicity, or his family members.
As a consequence, the leakage of this information may have serious implications if misused, as in genetic discrimination (e.g., for insurance, employment, or education) or blackmail~\cite{gottlieb2001us}. A true story exemplifying genetic discrimination was shared by Dr. Noralane Lindor at the Mayo Clinic's Individualizing Medicine Conference (2012)~\cite{storyvideo}. During her study of a cancer patient, Dr. Lindor also sequenced the grandchildren of her patient, two of whom turned out to have the mutation for the same type of cancer\footnote{Having a genetic mutation for a cancer only probabilistically increases the predisposition to the cancer.}. One of these grandchildren applied to the U.S. army to become a helicopter pilot. Even though genetic testing is not a required procedure for military recruitment, as soon as she revealed that she previously went through the aforementioned genetic test, she was rejected for the position (in this case legislation does not apply to military recruitment, as will be discussed below).

%As mentioned, genomic data reveal information about the family members of an individual.
Ironically, the familial aspect of genomics complicates the problems revolving around
%genome
privacy. A recent example
%of this problem
is the debate between the family members of Henrietta Lacks and the medical researchers~\cite{NYT}. Ms. Lacks (deceased in 1951) was diagnosed with cervical cancer and some of her cancer cells were removed  for  medical  research. These cells later paved the way to important developments in medical treatment. Recently, researchers sequenced and published Ms. Lacks's genome without asking for the consent of her living family members. These relatives learned this information from the author of the bestselling book ``The Immortal Life of Henrietta Lacks''~\cite{skloot2010immortal}, and they expressed the concern that the sequence contained information about her family members. After complaints, the researchers took her genomic data down from public databases. However, the privacy-sensitive genomic information of the members of the Lacks family was already compromised because some of the data had already been downloaded and many investigators had previously published parts of the cells' sequence.  Although the NIH entered into an agreement with the Lacks family to give them a voice in the use of these cells~\cite{HuffPost}, there is no consensus about the scope of control that individuals and their families ought to have over the downstream of their cells. Thousands of people, including James Watson~\cite{nyholt2008jim}, have placed their genomic data on the Web without seeking permission of their relatives.

%Today, the most serious threat that can emerge from
One of the often voiced concerns regarding genomic data is its potential for discrimination. While, today, certain genome-disease and genome-trait associations are known, we do not know what will be inferred from one's genomic data in the future. In fact, a grandson of Henrietta Lacks expressed his concern about the public availability of his grandmother's genome by saying that ``the main issue was the privacy concern and what information in the future might be revealed''. Therefore, it is likely that the privacy-sensitivity of genomic data, and thus the potential threats will increase over time.

Threats emerging from genomic data are only possible via the leakage of such data, and, in today's healthcare system, there are several candidates for the source of this leakage. Genomic data can be leaked through a reckless clinician, the IT of a hospital (e.g., through a breach of the information security), or the sequencing facility. If the storage of such data is outsourced to a third party, data can also be leaked from such a database through a hacker's activity or a disgruntled employee. Similarly, if the genomic data is stored by the individual himself (e.g., on his smartphone), it can be leaked due to a malware. Furthermore, surprisingly, sometimes the leakage is performed by the genome owner. For example, on a genome-sharing website, openSNP\footnote{Hosted at \url{http://www.openSNP.org}} \cite{greshake2014opensnp}, people upload the variants in their genomes
%(e.g., genotype provided by 23andme.com)
-- sometimes with their identifying material, including their real names.
%Even though such people may think they have nothing to hide, they should also consider their family members (as discussed in Henrietta Lacks example) when sharing such data.

One way of protecting the privacy of individuals' genomic data is through the law or policy. In 2007, the U.S. adopted the Genetic Information Nondiscrimination Act (GINA), which prohibits certain types of discrimination in access to health insurance and employment. Similarly, the U.S. Presidential report on genome privacy~\cite{Bioethics} discusses policies and techniques to protect the privacy of genomic data. In 2008, the Council of Europe adopted the convention concerning genetic testing for health purposes~\cite{Euro-convention}. There are, in fact, hundreds of legal systems in the world, ranging in scope from federal to state / province, and municipality level and each can adopt different definitions, rights, and responsibilities for an individual's privacy.  Yet, while such legislation may be put into practice, it is challenging to enforce because the uses of data cannot always be detected.  Additionally, legal regimes may be constructed such that they are subject to interpretation or leave loopholes in place.
%and they do not cover the whole system.
For example, GINA does not apply to life insurance or the military~\cite{altman2002}. Therefore, legislation alone, while critical in shaping the norms of society, is insufficient to prevent privacy violations.

The  idea  of  using  technical  solutions  to  guarantee  the privacy  of  such  sensitive  and  valuable  data  brings  about interesting debates. On one hand, the potential importance of genomic data for mankind is tremendous. Yet, privacy-enhancing technologies may be considered as  an  obstacle to achieving these goals. Technological solutions for genome privacy can be achieved by various techniques, such as cryptography or obfuscation (proposed solutions are discussed in detail in Section~\ref{sec:solutions}). Yet, cryptographic techniques typically reduce the efficiency of the algorithms, introducing more computational overload, while preventing the users of such data from ``viewing'' the data. And, obfuscation-based methods
%on the other hand,
reduce the accuracy (or utility) of genomic data. Therefore, especially when human life is at stake, the applicability of such privacy-enhancing techniques for genomic data is questionable.

On the other hand, to expedite advances in personalized medicine, genome-phenome association studies
often require the participation of a large number of research participants. To encourage individuals to enroll in such studies, it is crucial to adhere to
%good
ethical principles, such as autonomy, reciprocity and trust more generally (e.g., guarantee that genomic data will not be misused). Considering today's legal systems, the most reliable way to provide such trust pledges may be to use privacy-enhancing technologies for the management of genomic data. It would severely discredit a medical institution's reputation if it failed to fulfill the trust requirements for the participants of a medical study. More importantly, a violation of trust could slow down genomic research (e.g., by causing individuals to think twice before they participate in a medical study) possibly more than the overload introduced due to privacy-enhancing technologies. Similarly, in law enforcement, genomic data (now being used in FBI's Combined DNA Index System -- CODIS) should be managed in a privacy-preserving way to avoid potential future problems (e.g., mistrials, law suits). %\textbf{Brad: I'm having a difficult time again. I am failing to see how we can just make a quick switch from privacy in medical research to privacy in a law enforcement system.  Maybe I'm missing what the big picture of this paragraph is.} \textbf{Erman: I modified it a bit. The big picture is if the privacy of genomic data is not protected in these domains (medical, law enforcement), the consequences might be more costly than the potential (extra) investment on privacy enhancing technologies, what do you think?} JP: Yes, I think the issue is now solved.

In short, we need techniques that will guarantee the security and privacy of genomic data, without significantly degrading the efficiency of the use of genomic data in research and healthcare. Obviously, achieving all of the aforementioned properties would require some compromise. Our preliminary assessment of expert opinion (discussed in Section~\ref{sec:survey}) begins to investigate what tradeoffs users of such data would consider appropriate.
%are willing to trade money and time for the privacy of genomic data, while they strongly disagree in compromising accuracy (or utility) at all. In other words, they may be willing to invest more to deploy new technologies, and they may be willing to wait longer to get the results of genomic tests for security and privacy.
%As discussed, legislation, policies, and technological solutions have their own drawbacks. Therefore, at this point, it is crucial to come up with hybrid techniques via the collaboration of researchers from different disciplines in order to let genomics and privacy walk hand-in-hand.

%\cite{survey2013, Cristofaro13}
%
%\cite{Greenbaum2011}

%\cite{survey2013, Cristofaro13, lunshof2008genetic, greely2007uneasy}

\section{Genomics/Genetics Expert Opinion}
\label{sec:survey}

%\textbf{JP: As observed by Brad, the paragraphs following each of the ``beliefs'' are partially redundant with statements already made before. It thus makes sense to remove them. I left them temporarily because I responded to some of Brad's observations. This is just for the sake of the discussion.}
%Up to this point, this SoK has provided insight into what genomic data is, what it is used for, and why privacy is relevant in this setting.
%As alluded to, there are many domains in which genomic data will be used and, while there have been surveys with potential research participants regarding their concerns and desire for privacy, to the best of our knowledge there have been no investigations that consider the perspective of the researchers themselves.

\subsection{Objective}
We explored the views of an opportunistically ascertained group of
biomedical researchers in order to probe levels of concern about privacy and
security to be addressed in formulating guidelines and in future research.

\subsection{Survey Design}
The field of genomics is relatively young, and its privacy implications are
still being refined. Based on informal discussions (primarily with computer
scientists) and our review of the literature, we designed a survey to learn
more about biomedical researchers' level of concern about genomics and privacy.
Specifically, the survey inquired about \emph{(i)} widely held assertions about genome
privacy, \emph{(ii)} ongoing and existing research directions on genome privacy, and
\emph{(iii)} sharing of an individual's genomic data, using the following probes in
Figure~\ref{fig:beliefstext}. The full survey instrument  is available at
http://goo.gl/forms/jwiyx2hqol. The Institutional Review Board (IRB) at the
University of Illinois at Urbana-Champaign granted an exemption for the survey
. Several prior surveys focused on genome privacy have been
conducted and have focused on the perspectives of the general
public~\cite{kaufman2009public,kaufman2012preferences,platt2013public,Cristofaro13} and geneticists~\cite{pulley2008attitudes}. Our survey is
different because it investigates the opinion of biomedical researchers with
respect to the intention of data protection by technical means.

\begin{figure}
\noindent\fbox{
	\centering
\begin{minipage}{0.95\linewidth}
{
\small
The eight probes used to explore opinions about genome privacy follows:
\begin{enumerate}
	\item Genome privacy is hopeless, because all of us leave biological
cells (hair, skin, droplets of saliva,...) wherever we go.
	\item Genomic data is not special and should be treated as any other
sensitive health data e.g. health record or mental health notes.
	\item Genome privacy is irrelevant, because genetics is non-deterministic
	\item Genome privacy should be left to bioinformaticians, they can provide
better privacy solutions than computer security, privacy and cryptography
community can.
	\item Genome privacy will be fully guaranteed by legislation
	\item Privacy Enhancing Technologies are a nuisance in the case of genetics:
genetic data should be made available online to everyone to facilitate
research, as done e.g. in the case of the Personal Genome Project
	\item Encrypting genomic data is superfluous because it is hard to identify a
person from her variants
	\item Advantages of genomic based healthcare justify the harm that genome
privacy breach can cause.
%\item Genome privacy is a lost cause because everyone leaves biological cells (hair, skin, droplets of saliva, ...) wherever they go. The genome can be sequenced from those cells.
%\item Genome privacy is irrelevant:  Genomic data is just an additional record in a patient's file and will be protected with existing methods data.
%\item Genome privacy is irrelevant, because genetics is non-deterministic.
%\item Genome privacy should be left to bio-informaticians, as they specialize in both biology and computer science.
%\item Genome privacy will be fully guaranteed by legislation.
%\item Privacy-enhancing technologies are a nuisance in the case of genetics: genomic data should be made available online to everyone to facilitate research, as done, for example, in the case of the Personal Genome Project.
%\item Confidentiality of genomic data (e.g., using encryption) is superfluous because it is hard to identify a person from her variants.
%\item The advantages of genomic-based healthcare justify the harm that genome privacy breaches can cause.
\end{enumerate}
}
\end{minipage}
}
\caption{Probes of attitudes}
\label{fig:beliefstext}
\end{figure}

\subsection{Data Collection Methodology}
We conducted our survey both online and by paper. Snowball
sampling~\cite{Goodman61} was used to recruit subjects for the online survey.
This approach enables us to get more responses, but the frame is unknown and
thus response rate cannot be reported. A URL for the online survey was sent to
the people working in genomics/genetics area (i.e., molecular biology
professors, bioinformaticians, physicians, genomics/genetics researchers) known
to the authors of this paper. Recipients were asked to forward it to other
biomedical experts they know. Email and Facebook private messages (as an easy
alternative for email) were used to conduct the survey.
%\textit{We only asked biomedical experts to fill out the survey and asked them
%to forward it only to biomedical experts.} A URL for the online survey was
%sent to the people working in genomics/genetics area  (i.e., molecular biology
%professors, bioinformaticians, physicians, genomics/genetics researchers)
%known to the authors of this paper and they were asked to forward it to
%similar people they know. Email and Facebook private messages (as an easy
%alternative for email) were used to conduct the survey. 
Eight surveys were collected by handing out paper copies to participants of a
genomics medicine conference. Overall the survey was administered to 61
individuals. 
%Assuming an arbitrarily large population of people familiar with genetics, 

\subsection{Potential Biases.}
We designed the survey to begin to explore the extent to which biomedical
researchers share concerns expressed by some computer scientists. While not
generalizable to all biomedical experts due to the method for recruiting
participants, their responses do provide preliminary insights into areas of
concern about privacy and security expressed by biomedical experts. More
research is needed to assess the representativeness of these views.

\subsection{Findings}
\begin{figure}
	\centering
	\includegraphics[width=1\textwidth]{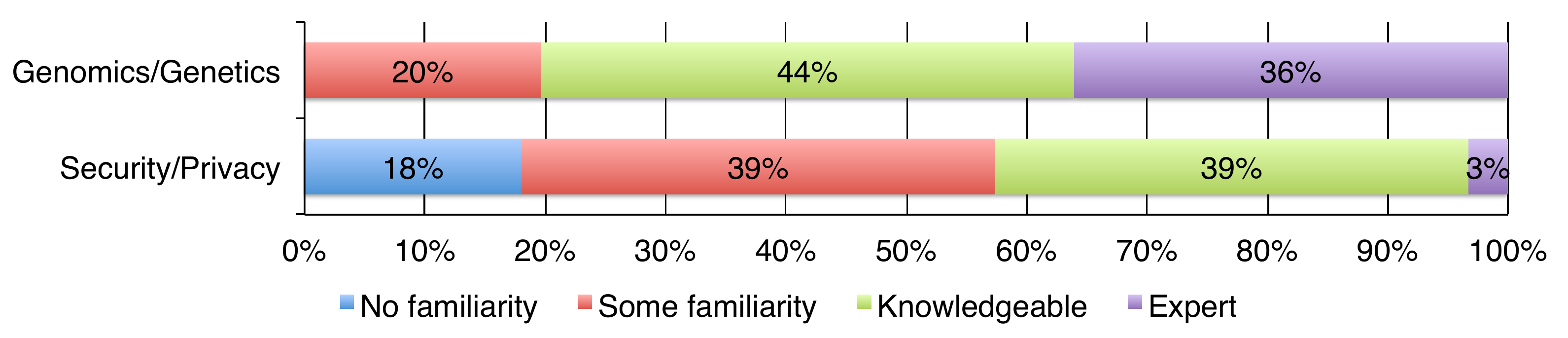}
	\caption{Self-identified expertise of the survey respondents}
	\label{fig:expertise}
\end{figure}

Approximately half of the participants were from the U.S. and slightly less
than half from Europe (the rest selected ``other''). The participants were
also asked to report their expertise in genomics/genetics and security/privacy.
We show these results in Figure~\ref{fig:expertise}.

\begin{figure}[h]
	\centering
	\includegraphics[width=1\textwidth]{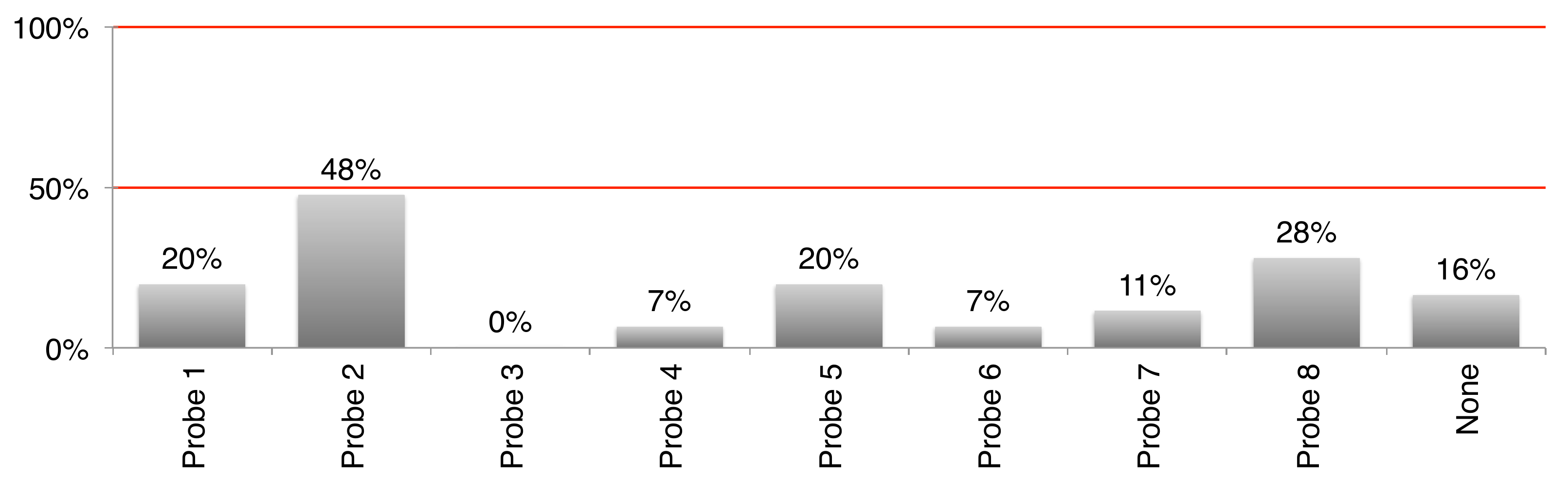}
	\caption{Response to the question: \emph{Do you believe that: (Multiple options can be checked)}. The probes are described in detail in Figure~\ref{fig:beliefstext}. ``None'' means the respondent does not agree with any of the probes.}
	\label{fig:beliefs}
\end{figure}

We asked whether the subjects agree with the statements listed in Figure 2
above. Figure~\ref{fig:beliefs} shows the results. 20\% of the respondents believe that
protecting genome privacy is impossible as an individual genomic data can be
obtained from his leftover cells (Probe 1). Almost half of the respondents
consider genomic data to be no different than other health data (Probe 2). Even
though genomic information is, in most instances, non-deterministic, all
respondents believe that this fact does not reduce the importance of genome
privacy (Probe 3).  Only 7\% of our respondents think that protecting genome
privacy should be left to bioinformaticians (Probe 4). Furthermore, 20\% of the
respondents believe that genome privacy can be fully guaranteed by legislation
(Probe 5). Notably, only 7\% of the respondents think that privacy enhancing
technologies are a nuisance in the case of genetics (Probe 6). According to
only about 10\% of the respondents, the confidentiality of genomic data is
superfluous because it is hard to identify a person from her variants (Probe
7). And, finally, about 30\% of the respondents think that advantages that will
be brought by genomics in healthcare will justify the harm that might be caused
by privacy issues (Probe 8).

%Overall we conclude that the eight beliefs we informally
%collected do not reflect the opinion of the biomedical community.

%We now comment on our findings from the survey. 
%We now present results of the survey.\\

\begin{figure}
	\centering			
	\includegraphics[width=1\textwidth]{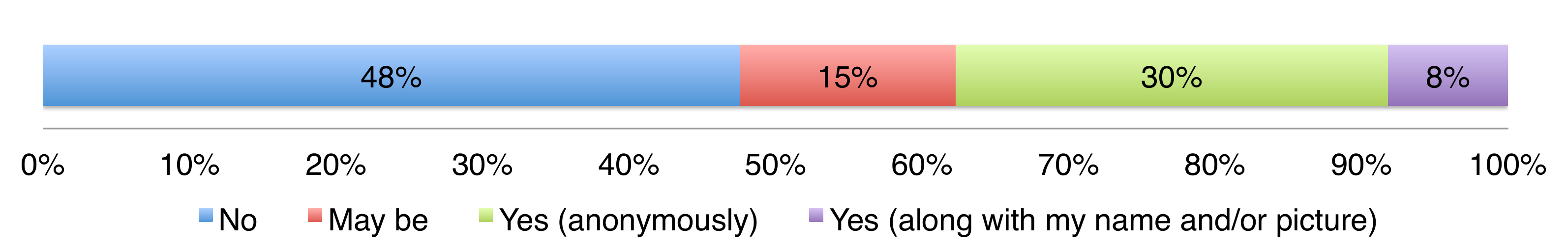}
	\caption{Response to the question: \emph{Would you publicly share your
		genome on the Web?}}
	\label{fig:sharing}
\end{figure}

\begin{figure}
	\centering
	\includegraphics[width=1\textwidth]{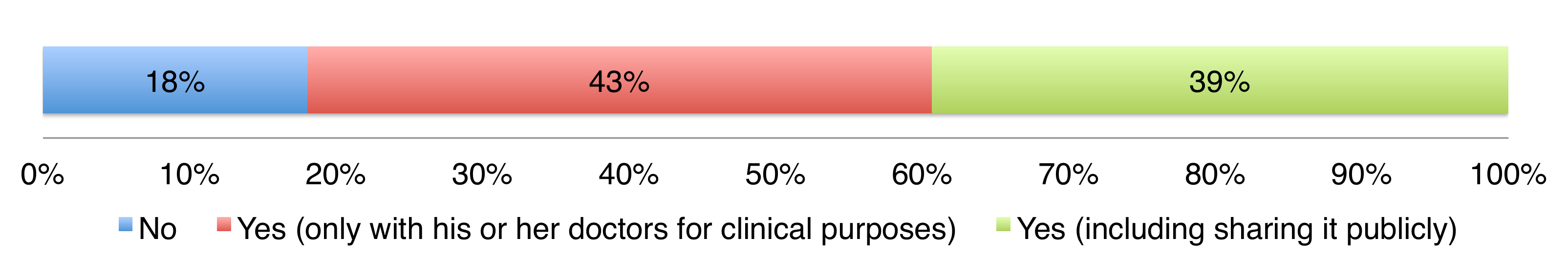}
	\caption{Response to the question: \emph{Assuming that one's genomic data
		leaks a lot of private information about his or her relatives, do you think
			one should have the right to share his or her genomic data?}}
	\label{fig:autonomy}
\end{figure}

%The importance of genome
%privacy has been advertised in several research papers and articles
%(e.g.,~\cite{survey2013}). 
%Note: I have cited this paper at another place.
We asked participants whether they would share their genomes on the Web
(Figure~\ref{fig:sharing}). $48\%$ of the respondents are \emph{not} in favor
of doing so, while $30\%$ would reveal their genome anonymously, and $8\%$
would reveal their identities alongside their genome. We also asked
respondents how they think about the scope of the individual's right to share
one's genomic data given that the data contain information about the
individual's blood relatives. Figure~\ref{fig:autonomy} shows that only $18\%$
of the respondents think that one should \emph{not} be allowed to share, $43\%$ of
the respondents think that one should be allowed to share only for medical
purpose, and $39\%$ of the respondents think that one should have the right to
share her genomic data publicly.  

%% $30\%$ of the subjects think that anonymization (or
%% pseudonymization) is sufficient to hide their identity. Recently, Humbert et.
%% al.~\cite{humbert2013} showed that the privacy loss of an individual due to
%% publicly available genomes of his family members is not negligible (discussed
%% in detail in Section~\ref{sec:riskassess}). Nevertheless, we observe that
%% $39\%$ support the right to publish one's genome (Figure~\ref{fig:autonomy}).
%% %\todo{Are they aware of the recent findings?}

\begin{figure}[h]
	\centering
	\includegraphics[width=1\textwidth]{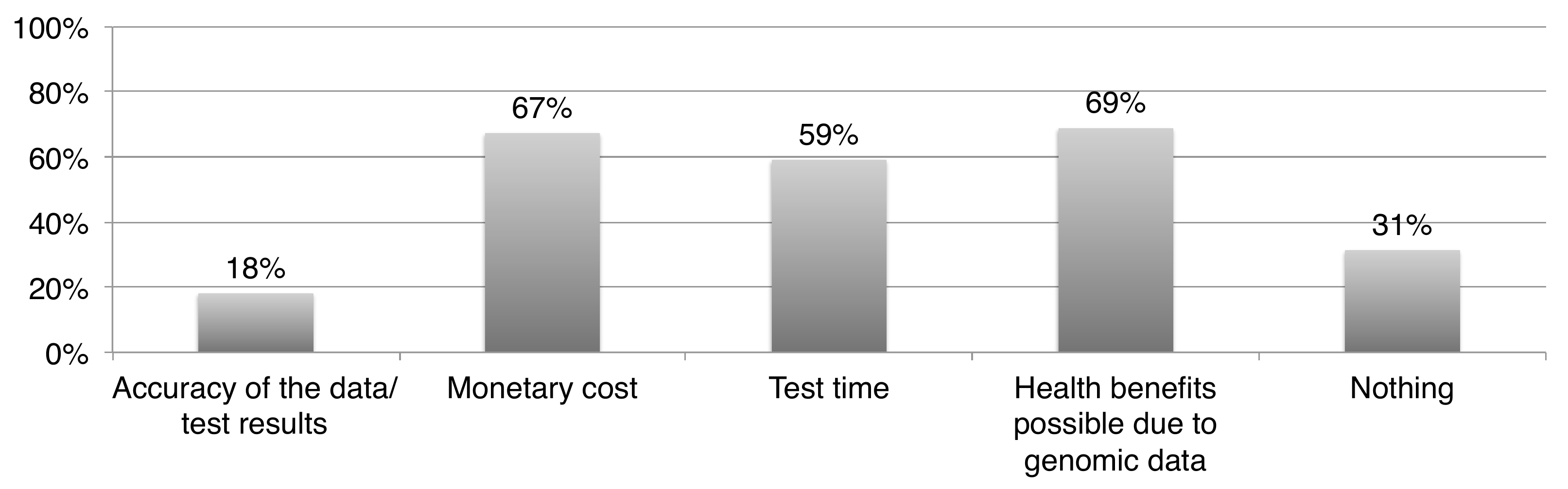}
	\caption{Response to the question: \emph{What can we compromise to improve privacy of genomic data? (Multiple options can be checked)}}
	\label{fig:compromise}
\end{figure}

As discussed in Section~\ref{sec:importance}, there is a tension between the
desire for genome privacy and biomedical research. Thus, we asked the survey
participants what they would trade for privacy. The results (shown in
Figure~\ref{fig:compromise}) indicate that the respondents are willing to trade
money and test time (duration) to protect privacy, but they usually do not
accept trading accuracy or utility.\\

\begin{figure}[h]
	\centering
	\includegraphics[width=1\textwidth]{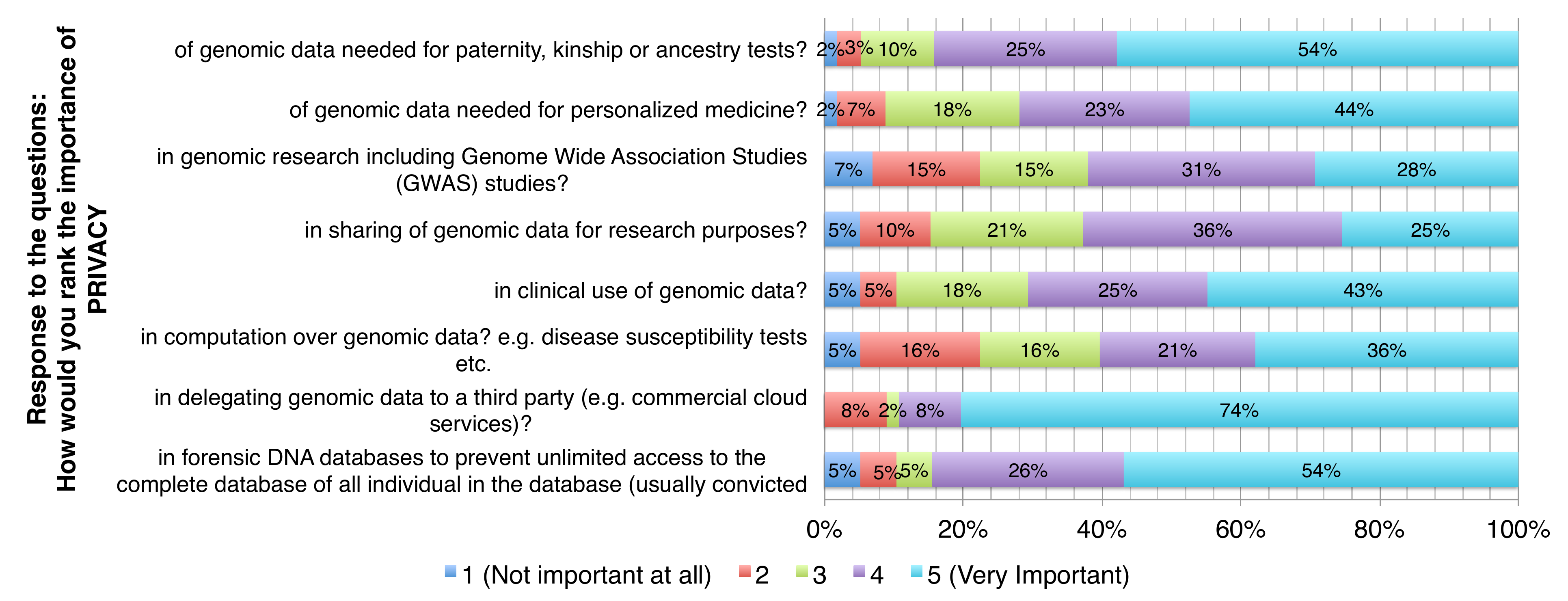}
	\caption{Relevance of genome privacy research done by the computer science
	community.}
	\label{fig:privacyquestios}
\end{figure}

We also asked the respondents to evaluate the importance of existing and
ongoing research directions on genome privacy (as discussed in detail in
Section~\ref{sec:solutions}), considering the types of problems they are trying
to solve (Figure~\ref{fig:privacyquestios}). The majority of respondents think that
genomics privacy is important.
%in all domains (i.e., healthcare, research, recreational
%genomics, legal and forensics) and think that privacy is the most important
%$74\%$ of subjects consider privacy in delegating genomic data to a third party (e.g., public cloud) as ``very important'', while $0\%s$ think that it is ``Not important at all''. 
%The subjects
%think that privacy is the less important when genomic data is used for
%research or healthcare issues.

\subsection{Discussion}
Our results show that these biomedical researchers believe that genomic
privacy is important and needs special attention. 
Figure~\ref{fig:beliefs} shows that, except for Probes 2 and 8, 80\% of the
biomedical experts do not endorse the statements listed in
Figure~\ref{fig:beliefstext}. Approximately three-quarters of the biomedical
experts believe that advantages of genome-based healthcare do not justify the
harm that can be caused by the genome privacy breach. Probe 2 is an
interesting result in that about half of the biomedical experts believe that
genomic data should be treated as any other sensitive health data. This seems
reasonable because, at the moment, health data can, in many instances, be more
sensitive than genomic data. The biomedical community also agrees on the
importance of current genome privacy research. Figure~\ref{fig:privacyquestios}
shows that these biomedical researchers ranks the placement of genomic data to the
cloud as their prime concern. Moreover, they agree with the importance of other
genome privacy research topics shown in Figure~\ref{fig:privacyquestios}.

We provide additional results stratified according to the expertise of the
participants in the appendix.

\section{Known Privacy Risks}
\label{sec:riskassess}

%What do we now know about the threats to the privacy of genomic data? We overview studies in a wide range of areas.

%\revised{In this section, we survey a wide spectrum of privacy threats to human genomic data, as reported by prior research.}

In this section, we survey a wide spectrum of privacy threats to human genomic data, as reported by prior research.

%\TODO{\cite{lowrance2007ethics,lunshof2008genetic,greely2007uneasy,ramos2013mechanism}}

\subsection{Re-identification Threats}
\label{sec:reid}

\textit{Re-identification} is probably the most extensively studied privacy risk in dissemination and analysis of human genomic data. In such an attack, an unauthorized party looks at the published human genomes that are already under certain protection to hide the identity information of their donors (e.g., patients), and tries to recover the identities of the individuals involved. Such an attack, once it succeeds, can cause serious damage to those donors, e.g., discrimination and financial loss.  In this section, we review the weaknesses within existing privacy protection techniques that make this type of attack possible.

\paragraph{\textbf{Pseudo-anonymized Data}} A widely used method for protecting health information is the removal of explicit and quasi-identifying attributes (e.g., name and date of birth). Such redaction meets legal requirements to protect privacy (e.g., de-identification under the U.S. Health Insurance Portability and Accountability Act) for traditional health records. However, genomic
data cannot be anonymized by just removing the identifying information. There is always a risk for the adversary to infer the \textit{phenotype} of a DNA-material donor (that is, the person's observable characteristics like eye/hair/skin colors), which will lead to her identification, from her \textit{genotypes} (her genetic makeup). Even though the techniques for this purpose are still rudimentary, the rapid progress in genomic research and technologies is quickly moving us toward that end. Moreover, re-identification can be achieved through inspecting the \textit{background} information that comes with publicized DNA sequences~\cite{gitschier2009inferential,gymrek2013identifying,Hayden2013}. As an example, genomic variants on the Y chromosome have been correlated with surnames (for males), which can be found out using
public geneology databases. Other instances include identifying Personal Genome Project (PGP) participants through public demographic data~\cite{sweeney2013unpub}, recovering the identities of family members from the data released by the 1000 Genome Project using public information (e.g. death notices)~\cite{Malin2006}, and other correlation attacks~\cite{MalinS04}. It has been shown that even cryptographically secure protocols leaks a lot of information when used for genomic data~\cite{goodrich2009mastermind}.

\paragraph{\textbf{Attacks on Machine Learning Models}}
Most attacks on genomic data use the entire dataset for the attack. Recently,~\cite{fredrikson2014privacy} showed that even the machine learning models trained on the genomic data can reveal information about the people whose data was used for training the model as well as any arbitrary person given some background information.
%They study Warfarin dosage example in detail and show that the attack has significant advantage compared to a random guess. 
%The authors further show that differential privacy is not useful for Warfarin dosage machine learning model.

\subsection{Phenotype Inference}

Another critical privacy threat to human genome data is inference of sensitive phenotype information from the DNA sequence.  Here we summarize related prior studies.

\paragraph{\textbf{Aggregate Genomic Data}} In addition to the aforementioned re-identification threats (discussed in Section~\ref{sec:reid}, which comes from the possible correlation between an individual's genomic data and other public information, the identity of a participant of a genomic study can also be revealed by a ``second sample'', that is, part of the DNA information from the individual. This happens, for example, when one obtains a small amount of genomic data from another individual, such as a small set of her single nucleotide polymorphisms (SNPs), and attempts to determine her presence in a clinical study on HIV (a phenotype), based on anonymized patient DNA data published online.  This turns out to be rather straightforward, given the uniqueness of individual's genome. Particularly,
in 2004, research shows that as few as 75 independent SNPs are enough to uniquely distinguish one individual from others~\cite{lin04}.  Based on this observation, the genomic researchers generally agree that such DNA raw data are too sensitive to release through online repositories (such as the NIH's PopSet
resources), without proper agreements in place. An alternative is to publish ``pooled'' data, in which summary
statistics are disclosed for the case and control groups of individuals in a study.

Yet,~\cite{homer2008resolving} showed that when an adversary had access to a known participant's genome sequence, they could determine if the participant was in a certain group.
Specifically, the researchers compared one individual's DNA sample to the rates at which her variants show up in various study populations (and a \textit{reference}
population that does not include the individual) and applied a statistical hypothesis test to determine the
likelihood of which group she is in (i.e., case or reference).
%they show a very high probability of success.
{\em The findings of the work led the NIH, as well as the Wellcome Trust in the UK, to
remove all publicly available aggregate genomic data from their websites}. Ever since,
researchers are required to sign a data use agreement (prohibiting re-identification) to access such
data~\cite{zerhouni2008protecting}, the process of which could take several months.
%Obtaining access to such aggregate data can take several months.
At the same time, such attacks were enhanced. First, Homer's test statistic was improved through exploitation of
%the logarithm of
genotype frequencies~\cite{jacobs2009new}, while
%and an individual genotype
%can improve the predictive power of Homer's attack~\cite{jacobs2009new}.
an alternative,
based on linear regression was developed to facilitate more robust inference
attacks~\cite{masca2011participant}.  Wang et. al.~\cite{wang2009learning} demonstrated, perhaps, an even more powerful attack by
showing that an individual can be identified even from the aggregate
statistical data 
%(coefficient of determination --$r^2$-- values)
(linkage disequilibrium measures) published in
research papers. While the methodology introduced in
\cite{homer2008resolving} requires on the order of 10,000 genetic variations (of the target individual), this new attack requires only on the order of 200. Their approach even shows the possibility of recovering part of the DNA raw sequences for the participants of biomedical studies, using the statistics including $p$-values and coefficient of determination ($r^2$) values.

Quantification of information content in aggregate statistics obtained as an output of genome-wide association studies (GWAS)
shows that an individual's participation in the study and her phenotype can be
inferred with high accuracy~\cite{im2012sharing,craig2011assessing}. Beyond these works, it has been shown that a Bayesian network could be
leveraged to incorporate additional background information,
and thus improve predictive power~\cite{clayton2010inferring}. It was recently shown that RNA
expression data can be linked to the identity of an individual through the inference of SNPs~\cite{schadt2012bayesian}.
%genotypes can be inferred only from the RNA expression data and can then be used to accurately identify individuals ~\cite{schadt2012bayesian}.

Yet, there is debate over the practicality of such attacks. Some researchers believe that individual identification from pooled data is hard in
practice~\cite{Braun2010,sankararaman2009genomic,visscher2009limits,Gilbert2008}. In particular, it has been shown that the assumptions required to accurately
identify individuals from aggregate genomic data rarely hold in
practice~\cite{Braun2010}. Such inference attacks depend upon the ancestry
of the participants, the absolute and relative number of people in case and
control groups, and the number of SNPs~\cite{masca2011participant} and the availability of the second sample. Thus, the
false positive rates are much higher in practice.
Still, others believe that publication of complete
genome wide aggregate results are dangerous for privacy of the participants~
\cite{lumley2010potential,church2009public}. Furthermore, the NIH continues to adhere to
its policy of data use agreements.

Beyond the sharing of aggregate data, it should be recognized that millions of people are sequenced or genotyped
for the state-of-the-art GWAS studies. This sequenced data is shared among
different institutions with inconsistent security and privacy
procedures~\cite{pmid23765454}. On the one hand, this could lead to serious backlash and fear to participate in such
studies. On the other hand, not sharing this data could severely impede
biomedical research. Thus, measures should
be taken to mitigate the negative outcomes of genomic data sharing~\cite{pmid23765454}.
%Participants are more restrictive in sharing their genomic data compared to normal data sharing in biomedical research~\cite{pmid22213783}.

\paragraph{\textbf{Correlation of Genomic Data}}
Partially available genomic data can be used to infer the unpublished genomic data due to
%so-called
linkage disequilibrium (LD), a correlation between regions of the genome~\cite{halperin2009snp,marchini2010genotype}. For example, Jim Watson
(the discoverer of DNA) donated his genome for research but concealed his ApoE gene, because it reveals susceptibility to Alzheimer's disease. Yet, it was shown
that the ApoE gene variant can be inferred from the published genome~\cite{nyholt2008jim}. Such completion attacks are quite relevant in
DTC environments, where customers have the option to hide some of the variants
related to a particular disease. 

While all the prior genomic privacy attacks
exploit low-order SNP correlations, \cite{samani2015quantifying} show that
high-order SNP correlations result in far more powerful attacks.

\cite{wagner2015metrics} investigates 22 different privacy metrics to study
which metrics are more meaningful to quantify the loss of genomic privacy due
to correlation of genomic data. 
%\textbf{Brad: Are you going to talk about the other attacks over the past
%several years, such as the work of Schadt, Cox, Ehrlich, and your own?}

\paragraph{\textbf{Kin Privacy Breach}}
A significant part of the population does not want to publicly release their genomic
data~\cite{mcguire2011share}. Disclosures of their relatives can thus threaten the privacy of such
people, who never release their genomic data.
% but their genomic information is inferred from their relatives who do.
The haplotypes of the individuals \textit{not} sequenced or genotyped can be obtained using LD-based completion
attacks~\cite{kong2008detection}. For instance, if both parents are genotyped, then most of
the variants for their offspring can be inferred.  The genomic data of family
members can also be inferred using data that has been publicly shared by blood
relatives and domain-specific knowledge about genomics~\cite{humbert2013}.
%They propose an algorithm to model this
Such reconstruction attacks %\textbf{Brad: This is the first mention of a
%``reconstruction. Perhaps you want to mention it earlier?}
can be carried out
using \textit{(i)} (partial) genomic data of a subset of family members,
%			\textit{(ii)} the correlation
%between the variables on the genome,
	and \textit{(ii)} publicly-known genomic background information (linkage
	disequilibrium and minor allele frequencies (MAFs).
	\ignore{\textbf{Brad: This Not sure what this is.  Are you talking about
	domain-specific knowledge again?}.}
%For the efficiency of such an algorithm, the authors represent this attack as
	%an inference problem (to infer the variants of the family members from the
	%available data). They, model the familial relationships, variants (in the
	%form of SNPs) on the DNA, and the correlation between the variants on a
	%factor graph. They then use a belief
%propagation algorithm~[1,2] \textbf{Brad: citation missing} to efficiently
	%infer the variants on the factor graph via message passing. The authors
	%also show how
This attack affects individuals whose relatives publicly share genomic data (obtained
using DTC services)
%like 23andMe)
on the Internet (e.g. on openSNP~\cite{greshake2014opensnp}).
%In particular, they deanoymized several
The family members of the individuals who publish their genomic data on
openSNP can be found on social media sites, such as Facebook~\cite{humbert2013}.

Note that ``correlation of genomic data'' and ``kin privacy breach'' attacks are based on different structural aspects of genomic data. While correlation attacks are based on the linkage disequilibrium (LD), which is a genetic variation within an individual's genome. A kin privacy breach is caused by genomic correlations among individuals. Moreover, a kin privacy breach can also be realized through phenotype information alone. For instance, a parent's skin color or height can be used to predict their child's skin color or height.
%Note that ``Correlation of genomic data'' and ``Kin privacy breach'' attacks are based on different structural aspects of the genomic data. While correlation attacks are based on the linkage disequilibrium (LD), which is a genetic variation within an individual's genome; kin privacy breach is caused by genomic correlations among individuals. Moreover, kin privacy breach can also occur due to phenotype information alone, for example, parents' skin color or height can be used to predict child's skin color or height.
 
\subsection{Other Threats}

In addition to the above threats, there are a few other genome-related privacy issues. We describe them below:

\paragraph{\textbf{Anonymous Paternity Breach}}
As mentioned above, the Y chromosome is inherited from father to son virtually intact and genealogy databases link this chromosome to the surname to model ancestry. Beyond the case discussed above, this information has been used to identify sperm donors in several cases.  For example, a 15 year boy who was conceived using donor sperm, successfully
found his biological father by sending his cheek swab to a genealogy service
and doing Internet search~\cite{Motluk2005,Stein2005}. Similarly, an adopted child was
able to find his real father with the help of a genealogy database (and substantial
manual effort)~\cite{Naik2009}. In short, DNA testing has made tracing anonymous
sperm donors easy and theoretically sperm donors can no more be
anonymous~\cite{Haupt2010}.

\paragraph{\textbf{Legal and Forensic}} DNA is collected for legal and forensic
purposes from criminals\footnote{\url{http://www.justice.gov/ag/advancing-justice-through-dna-technology-using-dna-solve-crimes}} and victims\footnote{\url{http://www.rainn.org/get-information/sexual-assault-recovery/rape-kit}}. On the one hand, forensic techniques are becoming more promising with the evolving technology~\cite{kayser2011improving,pakstis2010snps}.
%Specimen is collected from the victim to analyze DNA of the suspect, this sample will be contaminated by the victim's DNA.
On the other hand, abuse of DNA (e.g., to stage crime
scenes) have already baffled people and law enforcement
agencies~\cite{dirtysecret}. Some people like Madonna (the singer) are
paranoid enough about the misuse of their DNA that they hire DNA sterilization
teams to clean up their leftover DNA (e.g., stray hairs or saliva)~\cite{madonna}.   We are not aware of any privacy risk assessment
studies done primarily in legal and forensic context, in part because law
enforcement agencies store a very limited amount of genetic markers. Yet, in the
future, it could well happen that law enforcement agencies will have access to the
database of whole genome sequences. We discussed sperm donor paternity breach above which is also relevant in legal context.

%Find this \label{sec:risk_DTC}

%Humbert et. al.~\cite{humbert2013} bring a mathematical clarification to the
%case of Henrietta Lacks (discussed in Section~\ref{sec:importance}).  More
%specifically, they propose metrics to quantify personal and kin genomic (and
		%health) privacy. Furthermore, they show how the

%Users who share their
%genomic data on OpenSNP.org
%(a genome sharing website),
%found the family members of such users on Facebook, and showed the decrease in
%the genomic privacy of these family members (who did not share their genomic
%data) due to the genomic data shared on OpenSNP.org by their relatives.

%[1]  F. Kschischang, B. Frey, and H. A. Loeliger. Factor graphs and the
%sum-product algorithm. IEEE Transactions on Information Theory, 47, 2001.
%
%[2] J. Pearl. Probabilistic Reasoning in Intelligent Systems: Networks of
%Plausible Inference. Morgan Kaufmann Publishers, Inc., 1988.

\section{State-of-the-art Solutions}
\label{sec:solutions}

%\TODO{\cite{zhou2011esorics}}
%\TODO{\cite{Franz2013117}}
%
%\TODO{\cite{malin2011identifiability,church2009public}}
%
%Proposals: \cite{Malin05}

In this section, we provide an overview of technical approaches to address
various privacy and security issues related to genomic data. Despite the risks
associated with genomic data, we can find ways to mitigate them to move
forward~\cite{altman2013data}. Some solutions are efficient enough for
practical use, while others need further improvement to become practical.
In particular, practical solutions often exploit the special nature of the
genomic data to find ways to be efficient under relevant domain assumptions.

\subsection{Healthcare}
\paragraph{\textbf{Personalized medicine}} Personalized medicine promises to
revolutionize healthcare through treatments tailored to an individual's genomic
makeup and genome-based disease risk tests that can enable early diagnosis of
serious diseases.
%Patients are concerned about the privacy of their genomes, whereas healthcare
%organizations and pharmaceutical companies are worried about the privacy of
%their disease markers (a proprietary information of business importance).
Various players have different concerns here.  Patients, for instance, are
concerned about the  privacy of their genomes. Healthcare
organizations are concerned about their reputation and the trust of their
clients. And for-profit companies, such as pharmaceutical manufacturers, are
concerned about the secrecy of their disease markers (proprietary information
of business importance).

A disease risk test
%(or query)
can be expressed as a regular expression query taking
into account sequencing errors and other properties of sequenced genomic data.
Oblivious automata enable regular expression queries to be computed over genome sequence
data while preserving the privacy of both the queries and the genomic
data~\cite{troncoso2007privacy,frikken2009practical}. Cryptographic schemes
have been developed to delegate the intensive computation in such a scheme to a public
cloud in a privacy-preserving fashion~\cite{blanton2012secure}.

Alternatively, it has been shown that a cryptographic primitive called Authorized Private Set Intersection (A-PSI)
can be used in this setting~\cite{Baldi2011,de2012genodroid}. In
personalized medicine protocols based on A-PSI, the healthcare organization provides
cryptographically-authorized disease markers, while the patient supplies
her genome. In this setting, a regulatory authority, such as the U.S. Food and Drug Administration (FDA), can also certify
the disease markers before they can be used in a clinical setting.  Despite its potential, this
protocol has certain limitations.  First, it is not very efficient in terms of its
communication and computation costs.  Second, the model
assumes that patients store their own genomes, which is not necessarily the
case in practice.

To address the latter issue, it has been suggested that the storage of the homomorphically encrypted variants (e.g., SNPs) can be delegated to a
semi-honest third party~\cite{ayday2013ndss}. A healthcare organization can then request the third party to compute a disease susceptibility
test (weighted average of the risk associated with each variant) on the encrypted variants using an
interactive protocol involving \emph{(i)} the patient, \emph{(ii)} the healthcare organization and
\emph{(iii)} the third party. 
%In this protocol, the secret key is stored by the patient and to
%avoid sending the secret key directly to the healthcare organization, proxy
%re-encryption is applied and the secret key is divided into two shares. One share of the key
%is sent to the third party, while the other share is sent to the healthcare
%organization. After partially decrypting the test result, the third party
%sends the result to the healthcare organization, who fully decrypts the
%result using its share of the key. 
Additive homomorphic encryption enables a party with the public key to add
ciphertexts or multiply a plaintext constant to a ciphertext.
Additive homomorphic encryption based methods can also be used to conduct
privacy-preserving computation of disease risk based on both genomic and
non-genomic data (e.g., environmental and/or clinical data)~\cite{ayday2013healthtech}.
One of the problems with such protocols, however, is that storage of homomorphically encrypted variants
require orders of magnitude more memory than plaintext variants. However,
a trade-off between the storage cost and level of privacy can be composed~\cite{ayday2013globecom}.
%\textbf{This sentence needs to be tied in.  What's the method and how does it
%relate to what's in this paragraph?} \FIXED
%
A second problem is that when an adversary has knowledge of the LD between the genome regions
%variants SNPs
and the nature of the test, the privacy of the patients will decrease when tests are conducted on their homomorphically encrypted
variants. This loss of privacy can be quantified using an entropy-based metric~\cite{ayday2013wpes}.~\cite{danezis2014fast} propose two cryptographic protocols using framework proposed by~\cite{ayday2013wpes}. The first protocol involves a patient and a medical center (MC). MC encrypts the (secret) weights, sends them to the patient's smartcard, and operations are done inside the smartcard. This protocol also hides which and how many SNPs are tested. Second protocol is based on secret sharing in which the (secret) weights of a test is shared between the SPU and the MC. This protocol still relies on a smartcard (held by the patient) to finalize the computation. \cite{djatmiko2014secure} propose a secure evaluation algorithm to compute genomic tests that are based on a linear combination of genome data values. In their setting, a medical center prescribes a test and the client (patient) accesses a server via his mobile device to perform the test. The main goals are to \textit{(i)} keep the coefficients of the test (secret weights) secret from the client, \textit{(ii)} keep selection of the SNPs confidential from the client, and \textit{(iii)} keep SNPs of the client confidential from the server (server securely selects data from the client). They achieve these goals by using a combination of additive homomorphic encryption (Paillier's scheme) and private information retrieval. Test calculations are performed on the client's mobile device and the medical server can also perform some related computations. Eventually, client gets the result and shows it to his physician. As a case study, the authors implemented the Warfarin dosing algorithm as a representative example. They also implemented a prototype system in an Android App. In~\cite{karvelas2014privacy}, authors propose a technique to store genomic data in encrypted form, use an Oblivious RAM to access the desired data without leaking the access pattern, and finally run secure two-party computation protocol to privately compute the required function on the retrieved encrypted genomic data. The proposed construction includes two separate servers: cloud and proxy.
% A client can authorize a party, called “investigator” to perform a query on the data. Upon such a query, first, the investigator retrieves the required encrypted genetic data from the Oblivious RAM. Next, the cloud and the proxy run Yao’s garbled circuit protocol, with the inputs and circuit created by the investigator. Eventually, the output of the test is revealed only to the investigator.

%is useful to quantify such decrease in privacy~\cite{ayday2013wpes}.
%Non-genomic data (e.g., environmental and clinical data is as important as the genomic data (especially for early
%diagnosis).

Functional encryption allows a user to compute on encrypted data and learn the
result in plaintext in a non-interactive fashion. However, currently functional
encryption is very inefficient. \cite{Naveed2014CFE} propose a new
cryptographic model called ``Controlled Functional Encryption (C-FE)'' that
allows construction of realistic and efficient schemes. The authors propose
two C-FE constructions: one for inner-product functionality and other for any
polynomial-time computable functionality. The former is based on a careful
combination of CCA2 secure public-key encryption with secret sharing, while
the later is based on a careful combination of CCA2 secure public-key encryption
with Yao's garbled circuit. C-FE constructions are based on efficient
cryptographic primitives and perform very well in practical applications. The
authors evaluated C-FE constructions on personalized medicine, genomic patient
similarity, and paternity test applications and showed that C-FE provides much
better security and efficiency than prior work.

\paragraph{\textbf{Raw aligned genomic data}} Raw aligned genomic data, that is, the aligned outputs of a DNA sequencer, are often used
by geneticists in the research process.
%, Ayday et.
%al.~\cite{ayday2013esorics} propose a cryptographic technique.
Due to the limitations of current sequencing technology, it is often the case
that only a small number of nucleotides are read (from the sequencer) at
a time. A very large number of these \emph{short reads}\footnote{A short read corresponds to a sequence of nucleotides within a DNA molecule.
The raw genomic data of an individual consists of hundreds of millions of short reads. Each read typically consists of 100 nucleotides.}, 
%which cover the length of 
covering the entire genome are obtained, and are subsequently aligned, using
a reference genome. The position of the read relative to the reference genome
is determined by finding the approximate match on the reference genome. With today's sequencing techniques, the
size of such data can be up to 300GB per individual (in the clear), which makes
public key cryptography impractical for the management of such data.
%Therefore, the authors propose using
Symmetric stream cipher and order-preserving
encryption~\cite{agrawal2004order} 
%thus 
provide more efficient solutions for storing, retrieving, and processing
this large amount of data in a privacy-preserving way~\cite{ayday2013esorics}.
Order-preserving encryption keeps the ordering information in the
ciphertexts to enable range queries on the encrypted data. We emphasize that
order-preserving encryption may not be secure for most practical applications.

%It has been shown that symmetric cryptography can be used for privacy-preserving storage,
%retrieval and processing of raw genomic data~\cite{ayday2013wpes}.

%They suggest storing the raw data of the patients at a
%biobank. \textbf{Why biobank: Do you mean data management service?  A biobank
%is someplace that usually consists of freezers and laboratories.}
%\TODO{Rephrase} They outsource the management of cryptographic keys to a key
%manager instead of the owner of the genome (for the practicality of the
%proposed method). In their proposed framework, a healthcare provider asks for
%a definite range of nucleotides for one or more individuals from the biobank,
%who, in turn, provides all the short reads that include at least one
%nucleotide from the requested range.  Since each short read includes (about)
%100 nucleotides, it is possible that the healthcare provider may receive
%information that it did not ask for (or is not authorized for) from the
%biobank.  To avoid this violation, the authors propose an efficient masking
%technique using the properties of stream cipher before providing the
%corresponding short reads to the medical unit. They also show that significant
%leakage of privacy-sensitive information (in the raw genomic data) can occur
%unless the proposed masking technique is applied.
		
\paragraph{\textbf{Genetic compatibility testing}} Genetic compatibility
testing is of interest in both healthcare and
DTC settings. It enables a pair of individuals
to evaluate the risk of conceiving an unhealthy baby. In this setting,
PSI can be used to compute genetic compatibility, where one party submits the
fingerprint for his or her genome-based diseases, while the other party submits
her or his entire genome. In doing so, the couple learns their genetic
compatibility without revealing their entire genomes~\cite{Baldi2011}. This protocol
leaks information about an individual's disease risk status to the other party and
%\textbf{Brad: Is this a published protocol? If so, what's different than
%Baldi's?  If not, why are we reviewing it?} \FIXED \COMMENT{It is
%	\cite{Baldi2011} protocol, they have different protocols for pesonzlied
%medicine, genetic compability testing and paterntiy}
%The PSI-based genetic compatibility testing protocol of~\cite{Baldi2011} though practical
its requirements for computation and communication may make it impractical.
%\textbf{Brad: What is the protocol in question here?} \FIXED

%[6] R. Agrawal, J. Kiernan, R. Srikant, and Y. Xu, Order preserving
%encryption for numeric data, Proceedings of the 2004 ACM SIGMOD
%International Conference on Management of Data, pp. 563574, 2004.
%
%[7]  R. A. Popa, F. H. Li, and N. Zeldovich, An ideal-security protocol
%for order-preserving encoding, Proceedings of the 2013 IEEE Symposium on
%Security and Privacy, 2013.

\paragraph{\textbf{Pseudo-anyonymization}} Pseudo-anonymization is often performed by the healthcare organization
that collects the specimen (possibly by pathologists)
%or the place at which the specimen was collected
to remove patient identifiers before sending the specimen to a sequencing
laboratory. In lieu of such information, a pseudonym can be derived from the genome itself and public randomness,
%\textbf{Brad: What do you mean by ``public randomness''?} \COMMENT{Means
%randomness is not secret}
independently at the
healthcare organization and sequence laboratory for symmetric
encryption~\cite{cassa2013novel}. This process can mitigate sample mismatch at the
sequencing lab. However, since the key is derived from the data that is encrypted
using the same key, symmetric encryption should guarantee circular security
(security notion required when cipher is used to encrypt its own key),
an issue which is not addressed in the published protocol.
%~\cite{cassa2013novel}.
%\textbf{Brad: If we are going
%to use the phrase ``circular security'', it needs to be defined.} \FIXED

\paragraph{\textbf{Confidentiality against Brute-force Attacks}} History has
shown that encryption schemes have limited lifetime before they are broken.
Genomic data, however, has lifetime much longer than that of a state-of-the-art
encryption schemes. A brute-force attack works by decrypting the ciphertext
with all possible keys. Honey encryption~\cite{juels2014honey} guarantees that
a ciphertext decrypted with an incorrect key (as guessed by an adversary)
results in a plausible-looking yet incorrect plaintext. Therefore, HE gives
encrypted data an additional layer of protection by serving up fake data in
response to every incorrect guess of a cryptographic key or password. However,
HE relies on a highly accurate distribution-transforming encoder (DTE) over the
message space. Unfortunately, this requirement jeopardizes the practicality of
HE. To use HE the message space needs to be understood quantitatively, that is,
the precise probability of every possible message needs to be understood. When messages are not
uniformly distributed, characterizing and quantifying the distribution is
non-trivial. Building an efficient and precise DTE is the main challenge when
extending HE to a real use case;~\cite{huang2015genoguard} have designed such a
DTE for genomic data. We note that HE scheme for genomic data is not specific
to healthcare and is relevant for any use of genomic data.

\subsection{Research}
\label{subsol:research}

% is a popular method for learning genomic variants associated with a particular phenotype (e.g. asthma, cancer, heart rate, and mental illness)~\cite{gwas}.
%\textbf{Brad: Why do we jump directly into GWAS?  There many types of sequencing investigations that are based on exome, whole genome, etc.  This is why the NIH put a new proposed data sharing policy in September - http://grants.nih.gov/grants/guide/notice-files/NOT-OD-13-119.html.  Perhaps it would be bette to mention that there many ways for conducting genome-phenome associations and that GWAS is one?} \COMMENT{Brad can you help me address this comment}

\paragraph{\textbf{Genome-Wide Association Studies (GWAS)}} Genome-Wide
	Association Studies (GWAS), \footnote{\url{http://www.genome.gov/20019523}} are conducted by
analyzing the statistical correlation between the variants of a
\emph{case group} (i.e., phenotype positive)
and a \emph{control group} (i.e., phenotype negative). GWAS is one of the most common types of studies performed to learn genome-phenome associations.
In GWAS the aggregate statistics (e.g.,
$p$-values) are published in scientific articles and are made available to other
researchers. As mentioned earlier, such statistics can pose privacy threats as explained in
Section~\ref{sec:riskassess}.
%Protecting privacy of subjects in GWAS is of paramount importance~\cite{zerhouni2008protecting} for the progress of biomedical research.

%NIH used to make such statistics publicly available but they
%stopped after the Homer's attack~\cite{homer2008resolving}. Publishing such
%statistics is very useful for biomedical research and in general for human
%health, however privacy of the subjects involved in the study is of paramount
%importance~\cite{zerhouni2008protecting}.
%\textbf{Brad: I think that the
%previous statements can go away if we just refer to the earlier section where
%we talk about Homer's attack, right?} \COMMENT{Yes, I fixed it}

Recently, it has been suggested that such information can be protected through
the application of noise to the data.
In particular, differential privacy, a well-known
technique for answering statistical
queries in a privacy preserving manner~\cite{dwork2006differential}, was recently adapted 
%by Fienberg et.
%al.~\cite{fienberg2011icdmw} and Johnson et. al.~\cite{johnson2013kdd}
%proposed differentially
to compose privacy preserving query mechanisms for GWAS
settings~\cite{fienberg2011icdmw,johnson2013kdd}. A mechanism $K$ gives $\epsilon$-differential
privacy if for all databases $D$ and $D'$ differing on at most one record, the
probability of $K(D)$ is less than or equal to the probability of $\exp(\epsilon)
\times K(D')$. In simple words, if we compute a function on a database with
and without a single individual and the answer in both cases is
approximately the same, then we say that the function is differentially private.
Essentially, if the answer does not change when an individual is or is not in the
database, the answer does not compromise the privacy of that individual. \cite{fienberg2011icdmw} propose methods for releasing differentially private minor allele frequencies
(MAFs), chi-square statistics, $p$-values, the top-$k$ most relevant SNPs to a
specific phenotype, and specific correlations between particular pairs of
SNPs. These methods are notable because traditional differential privacy
techniques are unsuitable for GWAS due to the fact that the number of correlations
studied in GWAS is much larger than the number of people in the study.
%Thus, differential privacy mechanisms~\cite{fienberg2011icdmw},
However, differential privacy is typically based on a mechanism that adds noise
(e.g., by using Laplacian noise, geometric noise, or exponential mechanism),
and thus requires a very large number of research participants to guarantee
acceptable levels of privacy and utility. \cite{Yu2014133} have extended the work of~\cite{fienberg2011icdmw} to compute differentially private chi-square statistics for arbitrary number of cases and controls. \cite{johnson2013kdd} explain that computing the number of relevant SNPs and the pairs of correlated SNPs are the goals of a typical GWAS and are not known in advance. They provide a new exponential mechanism -- called a distance-score mechanism -- to add noise to the output. All relevant queries required by a typical GWAS are supported, including the number of SNPs associated with a disease and the locations of the most significant SNPs. Their empirical analysis suggests that the new approach produces acceptable privacy and utility for a typical GWAS.
%In~\cite{johnson2013kdd}
%%the latter paper
%, authors explain that computing the number of
%relevant SNPs and the pairs of correlated SNPs are the goal of a typical GWAS
%and are not known in advance. They provide a new exponential mechanism -- called distance-score mechanism -- to add noise to the output.  All
%relevant queries required by a typical GWAS are supported including the number
%of SNPs associated with a disease and locations of the most significant SNPs.
%Empirical analysis suggests that the new exponential mechanism based differentially
%private queries produce acceptable privacy and utility for a typical GWAS.

A meta-analysis of summary statistics from multiple independent cohorts is
required to find associations in a GWAS. Different teams of researchers often conduct studies
on different cohorts and are limited in their ability to share individual-level data due to
Institutional Review Board (IRB) restrictions. However, it is possible for the same participant to be in multiple studies, which
%Presence of duplicate individuals in multiple cohorts
can affect the results of a meta-analysis. It has been suggested that one-way
cryptographic hashing can be used to identify overlapping participants without
sharing individual-level data~\cite{turchin2012gencrypt}.

\cite{xie2014securema} proposed a cryptographic approach for privacy preserving genome-phenome studies. This approach enables privacy preserving computation of genome-phenome associations when the data is distributed among multiple sites. 

%A cryptographic approach to preserve privacy in GWAS was proposed in~\cite{xie2014securema}.
\paragraph{\textbf{Sequence comparison}} Sequence comparison is widely used in bioinformatics (e.g., in gene
finding, motif finding, and sequence alignment). Such comparison is computationally complex.
%\footnote{A problem size on the order of $10^{15}$\TODO{Detail}}  
Cryptographic tools such as fully homomorphic encryption (FHE) and secure
multiparty computation (SMC) can be used for privacy-preserving sequence comparison.
Fully homomorphic encryption enables any party with the public key to compute
any arbitrary function on the ciphertext without ever decrypting it. Multiparty
computation enables a group of parties to compute a function of their inputs
without revealing anything other than the output of the function to each other.
It has been shown that fully homomorphic encryption
(FHE), secure multiparty computation (SMC), and other traditional cryptographic
tools~\cite{atallah2003secure,Jha08} can be applied for comparison purposes, but they do not scale to a full human genome.
%, however, privacy issues hinder delegation of genomic data to cloud.
%As such, 
%the power of public cloud environments
%can be used
%for genomic sequence comparison purposes. 
Alternatively, more scalable provably secure
%Genomic data can be securely delegated to the cloud for
%sequence comparision with provaly secure cryptographic
protocols exploiting public clouds have been proposed~\cite{blanton2012secure,atallah2005secure}. 
Computation on the public data can be outsourced to a third party
environment (e.g., cloud provider) while computation on sensitive private
sections can be performed locally; thus, outsourcing most of the
computationally intensive work to the third party.  This computation
partitioning can be achieved using \textit{program specialization} which enables
concrete execution on public data and symbolic execution on the sensitive
data~\cite{wang2009ccs}. This protocol takes advantage of the fact that genomic computations
can be partitioned into computation on public data and private data,
exploiting the fact that 99.5\% of the genomes of any two individuals are
similar.

Moreover, genome sequences can be transformed into sets of offsets of different
nucleotides in the sequence to efficiently compute similarity scores (e.g., Smith-Waterman computations) on
outsourced distributed platforms (e.g., volunteer systems). Similar sequences
have similar offsets, which provides sufficient accuracy, and many-to-one
transformations provide privacy~\cite{szajda2006toward}.
%\textbf{Brad: First what? Later what?} \FIXED
Although this approach does not provide provable security, it does not leak
significant useful information about the original sequences.

Until this point, all sequence comparison methods we have discussed work on complete genomic
sequences. 
%As opposed to such techniques, 
Compressed DNA data (i.e., the variants) can be compared using
novel data structure called Privacy-Enhanced Invertible Bloom
Filter~\cite{eppstein2011privacy}. This method provides communication-efficient comparison schemes.

%\TODO{Merge Dynamic programming and sequence alignment}
\paragraph{\textbf{Person-level genome sequence records}} Person-level genome
sequence records contrast with the previous methods which obscure sequences and report on aggregated data rather than that of a single person.
%{\bf Sharing person-level genome sequences} is beneficial for biomedical research and in general for human health, but privacy implications prevent researchers from sharing the data.
Several techniques have been proposed for enabling privacy for person-level genome sequences.  For instance, SNPs from several genomic regions can be generalized into
more general concepts -- e.g.; transition (change of A$\leftrightarrow$G
or T$\leftrightarrow$C.), transversion (change of
A$\leftrightarrow$C, A$\leftrightarrow$T, C$\leftrightarrow$G or
G$\leftrightarrow$T.), and exact SNP positions
into approximate positions)~\cite{lin2002using}. This generalization makes re-identification of
an individual sequence difficult according to a prescribed level of protection. In particular,
%Tracking individual genomic data released from different institutions and linking it other information (e.g.; health record reveals sensitive information (including identify)).
$k$-anonymity can be used to generalize the genomic sequences such that a sequence is indistinguishable from
at least other $k-1$ sequences.  Also, the problem of SNP anonymization can be expanded to more complex variations of a genome using multiple sequence alignment and clustering methods
%such k-anonymous sequence sharing enables release of sequence data guaranteeing that it is ambiguous to at least one other sequence, thus preventing re-identification
~\cite{malinprotecting,LiWang12}. However, such methods are
limited in that they only work when there are a large number of sequences
with relatively small number of variations.

%\textbf{Brad: I'm a little worried that we are shifting between crypto and non-crypto methods.  It would be useful to provide a roadmap for presentation here: 1) Database queries, 2) Sharing data in the clear (obfuscation), and 3) Crypto computations?}
Given the limitations of generalization-based strategies, it has been suggested that cryptographic techniques might be more appropriate for maintaining data utility.  In particular, it has been shown that additive homomorphic encryption can be used to share encrypted data while still retaining the ability to
compute a limited set of queries (e.g., secure frequency count queries which are useful to many analytical methods for genomic data)~\cite{Kantarcioglu2008}.
Yet,
%~\cite{Kantarcioglu2008} protects the values of the SNPs,
this method leaks information in that it reveals the positions of the SNPs, which in turn reveals the type of test
being conducted on the data. Moreover, privacy in this protocol comes at a high cost of computation.
%\textbf{Brad: Do you want to talk about the paper we published on secure
	%coprocessors / tamper-resistent hardware for computations?  Seems relevant
		%here - http://www.ncbi.nlm.nih.gov/pubmed/22010157.}
%\COMMENT{Thanks, for the link, I have summarized the paper below.}

Cryptographic hardware at the remote site can be used as a trusted computation
base (TCB) to design a framework in which all person-level biomedical data is stored at a central remote server in encrypted form~\cite{canim2008}. The server can compute over
the genomic data from a large number of people in a privacy-preserving
fashion. This enables researchers to compute on shared data without sharing
person-level genomic data. This approach is efficient for typical biomedical
computations; however, it is limited in that trusted hardware tends to have relatively small memory capacities, which dictate the need for load balancing mechanisms.

%\textbf{Brad: Shouldn't this paragraph be moved above - where we talk about
	%short reads?  I understand that there's a distinction between healthcare
		%and research, but it seems that Wang's method is more appropriate here.
		%Also, this paragraph seems to provide the necessary information to
		%understand what's going on.}
%\COMMENT{I agree. I have moved the paragraph.}

\paragraph{\textbf{Sequence alignment}} Sequence alignment is fundamental to genome sequencing. The increase in the quantity of sequencing data is
growing at a faster rate than the decreasing cost of computational power, thus the delegation of
read mapping to the cloud can be very beneficial. However, such delegation
can have major privacy implications. Chen et.
al.~\cite{chen2012large} have shown that read mapping can be delegated to the
public cloud in a privacy preserving manner using a hybrid cloud based
approach. They exploit the fact that a sequence of a small number of
nucleotides ($\approx$20) is unique and two sequences of equal length with edit
distance of $x$, when divided into $x+1$ segments will have at least one matching
segment. Based on this fact, computation is divided into two parts: \emph{(i)} the public part is
delegated to the public cloud, in which the public cloud finds exact matches on encrypted
data and returns a small number of matches to the private cloud, whereas \emph{(ii)} the private
part takes place in a private cloud, which computes the edit distance using only the matches
returned by the public cloud.  This approach reduces the local computation by
a factor of 30 by delegating 98\% of the work to the public
cloud.

\subsection{Legal and Forensic}
\paragraph{\textbf{Paternity testing}} Paternity testing determines whether a certain male individual is the
father of another individual. It is based on the high similarity between the
genomes of a father and  child (99.9\%) in comparison to two unrelated
human beings (99.5\%). It is not known exactly which 0.5\% of the human genome
is different between two humans, but a properly chosen 1\% sample of the genome can
determine paternity with high accuracy~\cite{gibbs2006plos}. Participants may want to
compute the test without sharing any information about their genomes.

Once genomes of both individuals are sequenced, a privacy-preserving paternity
test can be carried out using PSI-Cardinality
(PSI-CA), where inputs to PSI-CA protocol are the sets of nucleotides comprising
the genome. The size of the human genome, or even 1\% of it, cannot be handled by current PSI and other SMC protocols.
However, by exploiting domain knowledge, the computation time can be reduced to 6.8ms and
network bandwidth usage to 6.5KB by emulating the Restriction Fragment Length
Polymorphism (RFLP) chemical test in software, which reduces the problem to
finding the intersection between two sets of size 25~\cite{Baldi2011}. Subsequent work
demonstrates a framework for conducing such tests on a Android smartphone~\cite{de2012genodroid}. Since the
ideal output of the privacy-preserving paternity test should be yes or no,
it cannot be obtained using custom PSI protocols, whereas generic garbled circuit
based protocols can be easily modified to add this capability~\cite{huang2011faster,huang2012private}. \cite{he2014identifying,hormozdiari2014privacy} propose cryptographic protocols for identifying blood relatives.

\paragraph{\textbf{Criminal forensics}} Criminal forensic rules enable law enforcement agencies to have unlimited access to the complete DNA record
database of millions of individuals, usually of convicted criminals (e.g.,
CODIS\footnote{Combined DNA Index System (CODIS), The Federal Bureau of
Investigation, \url{http://www.fbi.gov/about-us/lab/biometric-analysis/codis}} in the US). The motivation behind creating such a database
is to find a record
%in this database
that matches the DNA evidence from a crime
scene. Yet, providing unlimited access to law enforcement agencies is unnecessary
and may open the system to abuse.  Cryptographic approaches have been developed to
preserve the privacy of the records that fail to match the evidence from the
crime scene~\cite{bohannon2000cryptographic}. Specifically, DNA records can be encrypted using a key that depends upon certain
tests, such that when DNA is collected from a crime scene, the scheme will only allow decryption of the records that match the
evidence.

Finally, partial homomorphic encryption can be used for privacy-preserving matching of
Short Tandem Repeat (STR) DNA profiles in an honest-but-curious
model~\cite{bruekers2008privacy}. Such protocols (described in
Section~\ref{solssec:D2C}) are useful for identity, paternity, ancestry, and forensic
tests.

\subsection{Direct-to-consumer (DTC)}
\label{solssec:D2C}
Many DTC companies provide genealogy and
ancestry testing. Cryptographic schemes can be used to conduct these tests in
a privacy-preserving fashion. Partial homomorphic encryption can be cleverly used on STR profiles of
individuals to conduct \textit{(i)} common
ancestor testing based on the Y chromosome, \textit{(ii)} paternity test with one
parent, \textit{(iii)} paternity test with two parents, and \textit{(iv)} identity
testing~\cite{bruekers2008privacy}.

Despite the increasing use of
DTC genome applications, less focus has been given to the security and
privacy of this domain.
%and we believe research is needed to address issues in this area.
In particular, genomic data aggregation issues require special
attention because some companies 
%(e.g. OpenSNP.org) 
	allow people to publish
high-density SNP profiles online in combination with demographic and phenotypic data.
% along with their genome.

Focusing on kin genome privacy,~\cite{humbert2014reconciling} build a protection mechanism against the kinship attack~\cite{humbert2013} that uses DTC genomic data from openSNP~\cite{greshake2014opensnp}. The main focus of the work is to find a balance between the utility and privacy of genomic data. Every family member has a privacy constraint that he wants to protect. At the same time, some family members want to publish (part of) their genomes mainly to facilitate genomic research. The paper proposes a multi-dimensional optimization mechanism in which the privacy constraints of the family members are protected and at the same time the utility (amount of genomic data published by the family members) is maximized.  

\section{Challenges for Genome Privacy}
\label{sec:challenges}

%The new version begins here:

While the value of genome sequencing in routine care has yet to be fully
demonstrated, it is anticipated that the plummeting cost and
commoditization of these analyses will change the practice of medicine. Data
confidentiality and individual privacy will be central to the acceptance and
widespread usage of genomic information by healthcare systems. However, a
clear demonstration of the clinical usefulness of genomics is first needed for
doctors and other healthcare providers to fully embrace genomics
%and accept the constraints of genome
and privacy.

\subsection{Consumer-driven Genomics}
An
%additional,
unprecedented aspect of contemporary genomics that comes with its own set of issues for data
confidentiality is democratization, including facilitated access to
large-scale personal health-related data. Whereas medical and genetic
information used to be obtainable only through hospital or research
laboratories, people can now access their own genotyping or sequencing results
through direct-to-consumer companies such as 23andMe, as discussed before. On the research side, numerous participant-centric
initiatives have recently been launched (notably by citizens' networks such as
openSNP~\cite{greshake2014opensnp} and the Personal Genome Project). As a result,
%genotype/sequence
genomic
data are increasingly found outside the controlled cocoon of healthcare
systems or research. In particular, individual genetic results or aggregated
datasets are available on the Internet, often with non-existent or minimal
protection. On one hand, these crowd-based initiatives are very exciting,
because they have the potential to stimulate biomedical research, accelerate
discoveries and empower individuals.  Yet, on the other hand, they raise a
number of concerns about potential privacy risks (as highlighted in Section~\ref{sec:framework}). For example, privacy
risks must be assessed in the context of the extensive nature of information available on the Internet (including online social networks),
%, of course),
and not only within the narrower confines of genomic research or clinical care delivery.

\subsection{Privacy and the Benefits of Genomic Data}

\ignore{
Privacy protection is of utmost importance for medical data in general, but
even more so for genomic data. Although genomic data will (in most cases) be part of a patient's medical record in the future, as discussed in Section~\ref{sec:whyexceptional}, the data is different from most other medical data for several reasons: \textit{(i)} It can be used for predicting a number of severe diseases (including for the offspring); \textit{(ii)} it is highly stable in time; \textit{(iii)} it can be used for identification (and thus routinely used in forensics); \textit{(iv)} it can be obtained through direct-to-consumer genetic companies; and \textit{(v)} it is used for genealogical purposes (and in particular, it is an undisputable proof of paternity). The awareness of these characteristics has led legislators to set up regulations such as the Genetic Information Non-discrimination Act (GINA)~\cite{GINA} in the US and the convention concerning genetic testing for health purposes in Europe~\cite{Euro-convention}.
}

It is important to note that both a lack and an excess of privacy have the potential to derail the
expected benefits of genomics in healthcare and research. On one hand, the
efficient and secure handling of individual genotype and sequence data will
be central to the implementation of genomic medicine. The
%25-centuries old
Hippocratic Oath\footnote{See
\url{http://guides.library.jhu.edu/content.php?pid=23699&sid=190964} for a modern
version of the Hippocratic Oath.} remains a pillar of medical deontology and one of
the few stable concepts in the highly tumultuous history of medical practice.
The Hippocratic Oath contains a clear statement about the patient's privacy. Trust
is at the core of any successful healthcare system: any leakage of highly
sensitive genomic information may raise concerns and opposition in the
population and among political decision makers. Earning and conserving trust is
essential for hospitals and private companies that deal with genomics. As a
result, there
%seems to be
is a potential for a service industry securing genomic data, either by
providing ad hoc solutions or by fully supporting storage and delivery of
raw/interpreted sequence information. Fortunately, as detailed in
Section~\ref{sec:solutions}, there exist a variety of tools that can mitigate
the problem. 

On the other hand, an excess of privacy-related hurdles could slow
down research and interfere with large-scale adoption of genomics in clinical
practice. When designing privacy-preserving solutions for genomic data,
security and privacy researchers should keep in mind that most end-users are
not familiar with computer science and are almost exclusively interested in the
clinical utility of test results. %The securing of genomics should not be
%perceived as
%an additional burden by the medical and research communities: only then will
%such systems be widely adopted and used.\vspace{-5pt}
Education is again a fundamental requirement for privacy protection.
However, in bioinformatics curricula, students are trained to maximize the
information to be extracted from (biological) data. Usually, such curricula do
not address security and privacy concerns, because adversarial scenarios are
out of their scope. Conversely, computer scientists rarely have formal training
in biology, let alone genomics. But, they are trained in
security, notably because of the formidable challenge raised by the numerous
vulnerabilities of the Internet. Consequently, to properly address the concerns
about genomic data protection, there is a clear and strong potential of
cross-fertilization between these two disciplines.

\ignore{
Finally, as we have already mentioned the existence of legislation devoted to genomic data protection. Yet, such laws are difficult to enforce: If genomic data can be stealthily accessed, potential employers, bankers, and other decision makers will be tempted to make use of it (as recruiters do today by checking Facebook profiles of candidates). In addition, organized criminals (who are rarely deterred by laws) can misuse these data in multiple ways (blackmail, etc.).
}

\subsection{Acceptable Utility vs. Privacy of Genomic Data}
The balance between acceptable utility and privacy of genomic data needs to be considered
in context.

\paragraph{\textbf{Healthcare}} Patient-level information must be as precise as possible.
Because genomic data is used here to support clinical decision, including in
life-threatening situations, any decrease in data accuracy must be avoided.
Security of electronic medical records and other health-related information is
therefore most often guaranteed through restricted access (e.g., intranet use,
password and card identification) to unmodified data. It is important to note,
however, that genetic testing is typically not urgent, and that privacy-preserving
measures that would slightly delay a test result could be tolerated.
%\vspace{-3pt}

\paragraph{\textbf{Research}} Research on complex trait genomics relies on large datasets
on which genotyping or sequencing association studies can be run. To gain
statistical power and detect meaningful associations, it is often necessary to
merge many such studies through meta-analyses that can include
%genome-wide genotype
data from hundreds of thousands of individuals. Due to non-uniform use of
technological platforms, variation in time and place of genotyping, and
differences in analysis pipelines, some degree of noise is unavoidable.
An interesting avenue for research is here to empirically determine whether
differential privacy strategies (e.g., \cite{johnson2013kdd}) can be applied
without compromising discovery.
%\vspace{-3pt}

\paragraph{\textbf{Legal and forensics}} DNA collection and search for similarity pattern
in genomic data are used in criminal investigations and for other legal
purposes such as paternity testing. The accuracy of any test result is here again
an absolute requirement to avoid legal prejudice. Extremely stringent data
protection must also be ensured due to the highly sensitive nature of such
cases.

\paragraph{\textbf{Direct-to-consumer (DTC) genomics}} DTC companies providing individual genomic
data have a clear commercial incentive to protect customers' privacy, in order
to maintain trust and attract new customers. For example, the 23andMe webpage
states: \emph{``Your personalized web account provides secure and easy access
to your information, with multiple levels of encryption and security protocols
protecting your personal information.''}. Of course, these measures are
ineffective when individuals choose to unveil their full identity online
together with their genomic data, thereby putting their (and their blood
relatives) genome privacy at risk, either knowingly (as in the case of Personal
Genome Project participants) or out of naivety.

\section{Framework for Privacy-Preserving handling of Genomic data}
\label{sec:framework}

\begin{figure}[h]
\centering
%\hspace*{-1.5in}
\includegraphics[width=1\textwidth]{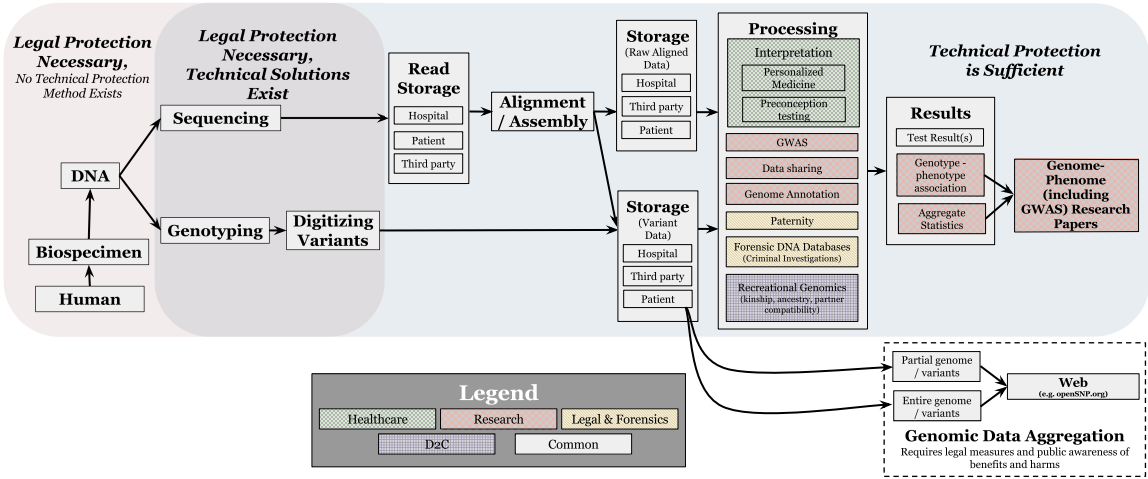}
%\makebox[\textwidth][c]{\includegraphics[width=1.4\textwidth]{Figures/frameworkfinal.eps}}
\caption{Genomic data handling framework: DNA is extracted from
	an individual's tissue or cells.  DNA is digitized either using sequencing
	(to obtain Whole Genome Sequence (WGS) or Whole Exome Sequence (WES)) or
	genotyping (to obtain variants, usually only SNPs). Reads obtained from
	sequencing are aligned to form the complete genome, while genotyped variants
	are digitized from the microchip array directly. Read data may be stored for
	later analysis.  The aligned genome can be either stored in raw form or
	compressed form (variations from a reference human genome). Medical tests
	and other types of computation shown in the figure can be performed either
	on raw aligned genome or just on variants. Possible outputs of computation
	are shown. Output depends on the type of computation and in some cases there
	is no output. The figure shows the genomic data aggregation problem caused
	by recreational genomics services. The figure is divided into three sections
	based on fundamental limitations of legal and technical measures for the
	protection of genomic data.  Legal protection is required for the left
	section, legal as well as technical protection is required for the middle
	section, while, in theory, technical solutions would suffice for the
	protection of the right section.  The legend shows which blocks are
	associated with different uses of genomic data. \textit{We use the word
		``patient'' in this paper to mean someone whose genome is sequenced or
genotyped and not necessarily a person who is ill.}}

\label{fig:flow}
\end{figure}

In this section, we provide a general framework for security and privacy in
the handling of genomic data. The framework is illustrated in
Figure~\ref{fig:flow}. As has been done throughout this paper, we divide the
framework into four categories: \emph{(i)} healthcare, \emph{(ii)} research,
\emph{(iii)} legal and forensics, and \emph{(iv)} direct-to-consumer genomics.
This classification is based on the most popular uses of genomic data;
however, we recognize that the boundaries between these categories are blurred
and there is significant overlap. For each of these we describe setting, threat
model, and solutions and open problems. The setting provides the most general
environment around the problem (e.g., we do not discuss the possibility of
outsourcing the computation as one can easily extend our setting to another
one involving a third party). 
%In this section, we always assume that the adversary is computationally bounded. We also assume that adversary can use all publicly available information (e.g., data from 1000 Genomes Project or public genealogy databases) to her advantage. Moreover, in some cases adversary might have access to private data. People can abuse their access to the private data, adversary can also steal the data, and data can be leaked from a lost laptop.
In this section, we assume that the adversary is computationally bounded. We further assume that the adversary can leverage all publicly available information (e.g., data from the 1000 Genomes Project or public genealogy databases) to her advantage. Moreover, in some cases, the adversary might have access to private data. For instance, people can abuse their access to private data, an adversary can also steal the data, and data can be extracted from a lost laptop.

\subsection{Biospecimen}
\label{ssec:biospecimen}
DNA is obtained in chemical form and then digitized. This
cyber-physical nature of DNA creates unique challenges for its protection.

\subsubsection{Threat Model}
\label{sssec:biothreatmodel}
In our threat model, the adversary is capable of \emph{(i)} obtaining DNA
from an individual's biological cells either voluntarily (e.g., for research
with informed consent) or involuntarily (e.g., leftover hairs or saliva on a
coffee cup), \emph{(ii)} sequencing or genotyping the DNA from
biospecimen, \emph{(iii)} interpreting the sequenced data to learn
identity, disease, kinship, and any other sensitive information, and
\emph{(iv)} linking the genomic data (or biospecimen) to the identity,
health record, or any arbitrary background information about the individual.
%We note that at least in the US one's abandoned biological cells (e.g.,
%saliva or hairs) can be \emph{legally} collected and used for any purpose.

\subsubsection{Solutions and Open Problems} Legal protection is
\textit{necessary} to protect the biospecimen and the DNA (in its chemical
form). \ignore{A simple solution to prevent the abuse of direct-to-consumer genomics
is to outsource biospecimen collection to hospitals or any other trustful
entity (like notary public) can guarantee that the specimen was collected and
enclosed in the presence of a credible individual.} However, a solution to this problem is
a subject of public policy and is outside the scope of this paper.

\subsection{Digitization: Sequencing/Genotyping}
\label{ssec:digitization}
%\textbf{Brad: should this be introduced in the Background or Introduction? It seems really late in the paper to introduce this now.}

\subsubsection{Setting} A biospecimen is obtained by an agency (e.g., hospital)
	and is sequenced or genotyped either by the same agency or by an external agency (e.g.,
	Illumina, 23andMe, etc.).

\subsubsection{Threat Model}
Since a biospecimen is required for digitization, we assume the threat
model discussed in Section~\ref{sssec:biothreatmodel} with the following
extensions: \emph{(i)} the adversary has full control over the entire sequencing
or genotyping infrastructure, \emph{(ii)} the adversary can be honest-but-curious
and can attempt to learn partial or entire genomic data or any information
derived from the genomic data, and \emph{(iii)} the adversary can be malicious and
can compromise the integrity of partial or entire genomic data.

\ignore{We consider the adversary (including
		sequencing/genotyping agency itself) with the following capabilities:
\begin{itemize}
	\item He can link a biospecimen, extracted DNA or sequenced genomic data to the identity of
	the subject/patient.
	\item He can be honest-but-curious and may want to learn partial or entire
		genomic data or any function of genomic data.
	\item He can be malicious or be compromised by an adversary and can compromise
	the integrity of partial or entire genomic data.
	\item He has full control over the entire sequencing infrastructure.
\end{itemize}
	}

\subsubsection{Solutions and Open Problems} Given the cyber-physical nature of DNA, it is not
possible to address this issue with technical measures alone. Both legal and
technical protections are required to protect against this threat. An external
agency cannot derive genomic data without a biospecimen, and legal
protection is required to prevent misuse. Sequencing machines are expensive and are manufactured by a limited number of companies. We envision a
well-regulated process for the manufacturing, procurement, and use of sequencing
machines and government regulations in place for it. The FDA
already regulates the manufacturing of medical devices\footnote{\url{http://www.fda.gov/medicaldevices/deviceregulationandguidance/postmarketrequirements/qualitysystemsregulations/}}~\cite{cheng2003medical}. Regular inspections would check for compliance. Under such
legal protections, sequencing machines could have tamper resistant trusted computing
base (TCB) that could output encrypted data
%as output of the machine,
such that the
sequencing agency could not access the plaintext genomic data.
%Digital rights management (DRM) based
%technologies could also be beneficial for sequencing machines.
%\COMMENT{Describe who the patient is (e.g., not a sick person, but someone who is sequenced.)}

\subsection{Storage}
%Genomic data can be stored by the patient herself, by a
%healthcare organization, or by a third-party (e.g., cloud service provider).
%, government entity etc.).
We assume that
once the adversary has access to the read data, it is easy to get raw aligned
data and variant data, hence we present storage of all three forms of data
together.
%in the following (in Figure~\ref{fig:flow}, we separate the storage of raw aligned data and variant data in order to clearly illustrate the data flow).

\subsubsection{Setting}
Genomic data can be stored by the \emph{(i)} patient\footnote{We use the word
	patient to mean an individual whose genome is sequenced or
genotyped and not necessarily a person who is ill.}, \emph{(ii)} healthcare
organization (e.g., as part of patient's EHR), or
\emph{(iii)} a third party.
%%We differentiate between the healthcare organization and third
%%party because patients trust healthcare providers with their sensitive
%%medical information, though this is not necessarily the case with a third
%%party.

\subsubsection{Threat Model}  For all settings, we assume that the lifetime of genomic data is much longer than the lifetime of a cryptographic algorithm. We consider the following threat models
%depending upon the setting:

\emph{Patient:} Storage media or the device storing genomic data can be lost,
	stolen, or temporarily accessed. A patient's computer can be attacked by an
	adversary (curious or malicious) to compromise confidentiality and/or
	integrity of the genomic data. We further assume that an adversary can
	associate the identity and
%arbitrary
background information (including
			phenotype information) from any arbitrary source. And, we
	assume that the adversary can use the compromised genomic data for any arbitrary
	purpose.\\
\indent \emph{Hospital:} We consider all of
	the threats described for the patient and the following additional threats. An insider (or hacker) has full access to the infrastructure
	and can link genomic information to the phenotypic information
	in the patient's health record.\footnote{We assume that data stored at the
		hospital is not anonymized.} We also consider the threat of the healthcare organization
		communicating incidental findings that are harmful -- violating \emph{the
			right not to know}.\footnote{
				For instance, a doctor telling a patient his increased
				susceptibility to Alzheimer's disease, when he does not want to know. We
					emphasize that defining what is harmful is an ethical issue and is
					out of scope of this study.}
					%For instance, this can occur when a doctor tells a patient that he has an increased
				%chance of getting Alzheimer's, when he does not want to know. We
				%	emphasize that defining what is harmful is an ethical issue and is
				%	out of scope of this study.
					We assume that the adversary can
					intentionally try to figure out variants of regions of the genome from variants in other regions,
					%which is a concern when it is the
					%case that
					e.g., to learn some sensitive SNPs that are removed to prevent incidental finding
					issues. We also assume that a healthcare organization could illegally collude with
					insurance agencies to facilitate discrimination based on genomic data.\\
\indent \emph{Third party:} We consider all of the threats discussed for the hospital
and following additional threat. The adversary, as the third party itself,
can be interested in de-anonymizing the anonymized genomic data or aggregate
genomic data.

\subsubsection{Solutions and Open Problems}
We report some solutions in Section~\ref{sec:solutions}. Users are generally
not equipped with skills and equipment to protect the security and privacy of
their genomic data. For the storage of genomic data, an option is to
store it on a cloud in an encrypted fashion, which makes the adversary's job
harder, as now it needs to circumvent cloud storage security measures and also
require to hack into user's computer to steal the decryption key. Efficient
encryption schemes allowing secure storage and computation are required.

\subsection{Alignment/Assembly}
As explained in Section~\ref{sec:solutions}, genomic data is obtained from the
specimen in the form of short reads. These short reads are then assembled using
alignment or assembly algorithms. Alignment is done by comparing the short
reads to the reference genome, and is computationally very intensive. Hence,
it can be economically beneficial to delegate the alignment to the cloud.

\subsubsection{Setting}
Short reads are obtained locally from the sequencing machine and alignment is delegated to an untrusted third
party.

\subsubsection{Threat Model}
We assume that the third party can be honest-but-curious, as well as
malicious, and can return incorrect results for economic or other malicious motives.

\subsubsection{Solutions and open problems}
We presented some solutions to this problem in Section~\ref{sec:solutions}~\cite{chen2012large}.
However, there are several problems with the most efficient solution to date. First, it is not provably
secure. Second, its security and efficiency requires that the read size
be greater than 100 nucleotides. Third, this scheme only works in a hybrid cloud environment and
requires local computation.
%JP's comment: the (commented) lines above are way too harsh.
Given that our opinion poll (described in Section~\ref{sec:survey}) shows that third party environments are of the greatest
concern to biomedical researchers, a provably secure and efficient solution that is capable
of aligning the human genome in a cloud computing environment is an important open research problem.

%At the same time, the integrity of the alignment process is critical and must be considered when designing such solutions.

\subsection{Interpretation}
Interpretation depends upon two private inputs:
the patient's genomic data and an interpretation algorithm (possibly from more than
one party). Given the complexity of genomic data, it is unlikely that any single party will have
a complete interpretation algorithm.  This makes the computation a multiparty
process between independent parties and the patient (or an agent of the patient, e.g., a hospital). Although each party
with a piece of the interpretation algorithm can compute independently with the
patient, collusion of any two parties may leak information about another
party's inputs. Moreover, the interpretation of one party may depend upon the
interpretation of another party. We assume that all of these parties can be
malicious and can collude to learn information (e.g., the patient's
genomic data or another parties' algorithm). In some cases, it is necessary to
sequence one's parent to draw conclusions, in which case
parents might also be concerned about their privacy.

\emph{Personalized medicine} is a special case of interpretation and depends
upon the patient's genomic data and disease markers (possibly distributed
among multiple parties).

\ignore{needs private inputs from two different parties, one
input being the patient's genomic data and other being the disease markers. We
assume that party responsible for the computation can be honest-but-curious,
in which case it is interested in learning any information about the patient's
genomic data and/or disease marker. It can also be maliciously corrupt in
which case it can cause to compute wrong results. We also assume that the adversary
can try to infer the disease markers by creating fake queries.
}

\emph{Preconception testing} is another special case of interpretation. It is different from personalized medicine because it is a
pre-pregnancy test and measures can be taken to conceive a healthy
child (as evident from www.counsyl.com success stories).
%As opposed to personalized medicine,
Additionally, the outcome of the preconception test almost \emph{always} depends
upon two people, each of whom might
prefer to not disclose their genomic status to the other.
%or any information inferred from it with each other.

\subsubsection{Setting} The computation is typically
performed on private data from multiple parties. The parties involved in the computation are
those
%parties
who input \textit{(i)} their genomic data and
%parties input
\textit{(ii)} interpretation
algorithms. The output of the computation should
only be released to the patient or authorized physician (possibly using the
infrastructure of a healthcare organization).

\subsubsection{Threat Model} We assume that all parties can be
honest-but-curious, malicious, or colluding (and possibly all at the same time).
They can use arbitrary background knowledge to breach privacy.
They may use falsified genomic data or a falsified interpretation algorithm an arbitrary number of times to ascertain another parties' private inputs. Furthermore, they may influence the results in an arbitrary manner.

\subsubsection{Solutions and Open Problems}
We discussed some of the solutions for the personalized medicine scenario
in Section~\ref{sec:solutions}. However, current solutions are limited to privacy-preserving disease
susceptibility tests. It is clear that computational solutions that support a broad range of computation over
genomic data are needed. At the same time, the design of such systems must be practical
%considered in design of such systems
and provide reasonable usability, accuracy, security, and privacy. 

%\textbf{Brad: What type of expectations are you talking
%about?} \FIXED

\subsection{Genome-Wide Association Studies (GWAS)}
\subsubsection{Setting} The genomic data from two groups of people are collected,
	one being the case group (i.e., people with the disease or some other trait) and the other being the control
	group (i.e., people without the disease). Statistical analysis is then conducted to discover
	the correlation between the disease and genetic variants.
% (or between different
%	variants).
The results
%statistics
are subsequently published in research papers and
%	detailed statistical results are
posted online possibly with restricted access (e.g., at dbGaP\footnote{\url{http://www.ncbi.nlm.nih.gov/gap}}).
%\textbf{Brad: Does this need to be repeated?}

\subsubsection{Threat Model}
An adversary may wish to determine if the victim is a GWAS participant or blood
relative of a GWAS participant. We assume that the adversary has access to the
high density SNP profile of the victim and also to a reference population
(which can be obtained from the same GWAS conducted on a different
 population). The attack succeeds if the adversary learns information
%additional advantage
from the data produced by GWAS, which she otherwise would not have learned.

\subsubsection{Solutions and Open Problems}
There are various solutions that could be applied in this setting.  We explained noise-based solutions, such as differential privacy, in
Section~\ref{subsol:research}. Yet, differential privacy-based solutions make data
more noisy, which make adoption of these approaches difficult. This is particularly problematic because
%In practice,
biomedical researchers and physicians
want more (not less) accurate data than is available today.
% due to physical and economical limitations.
An ideal solution should preserve the utility of data while preserving the privacy of
participants. We believe that more research is required in this
area to determine if noise-based approaches can lead to more usable and pragmatic data publications.
%realistic.
These approaches may, for instance, build upon well-established sets of practices from other communities.
%preventing sensitive information leakage from other forms of high-dimensional data.
For example, the Federal Committee on Statistical Methodology (FCSM)
has a long history of sharing information in a privacy-preserving manner.  These practices obey multi-level access principles
%such well-established set of practices. Census data is a good example of
%such data, which is controlled through multi-level anonymization.
and, to the best of the authors' knowledge, no significant privacy breach from
such domain has been reported.\\
\indent While the data disclosed by federal agencies is quite different from high-dimensional genomic data, it might be possible to
adapt these practices to balance the benefits and harms caused by public sharing of aggregate genomic data. These strategies may be composed of social and technical protections.  From a social perspective, a popular method to mitigate risk is through contractual agreements
%between research institutions (or investigators), such that
%they share this data
%with an
which prohibit the misuse of such data.
% agreement that privacy of the participants would be protected.
Such contracts could be complemented by cryptographic protocols that help preserve the privacy of
	the participants, particularly in settings in which the data is used in the secure
	computation and only the output of the computation is revealed to a specific
	party.

\subsection{Data sharing} The majority of genome-phenome discoveries come from very large
populations,
% which researchers were not able to observe otherwise. Large
%population of participants is crucial for GWAS and
sometimes on the order of millions of
participants.
%are required
%to draw conclusions from GWAS.
%Therefore,
Given the costs and scarcity of such resources, sharing data would fuel biomedical research.
%and thus for human health.
However, sharing this data entails privacy implications as discussed earlier.

It should be noted that individual-level genomic data from a large number of people is needed to conduct genome-phenome studies (e.g., GWAS). Moreover, the results of a typical GWAS are published as aggregate genomic data, which is usually made available to researchers at large (e.g., NIH requires results of all NIH-sponsored GWASs to be uploaded to dbGaP and published in research papers --- e.g., $p$-values of relevant SNPs). Therefore, we only focus on the sharing of individual level genomic data here.
%Note that individual level genomic data from a large number of people is needed to conduct genome-phenome studies (e.g., GWAS). Moreover, results of a typical GWAS are published as aggregate genomic data, which is usually made available to researchers at large (e.g., NIH requires results of all NIH sponsored GWASs to be uploaded to dpGaP\footnote{\url{http://www.ncbi.nlm.nih.gov/gap}}) and published in research papers (e.g., $p$-values of relevant SNPs). Therefore, we only focus on the sharing of individual level genomic data here.
%Privacy-preserving solutions are required to share this data without compromising the privacy of participants.

\subsubsection{Setting}
%Large number of patients are sequenced for research purposes.
Genomic data needs to be shared among different research institutions, possibly under different jurisdictions.
Privacy-preserving solutions can be built
in the following settings: \emph{(i)} all data delegated to and computation
done on a trusted party (e.g., a governmental entity), \emph{(ii)} all data
delegated to and computation done on an untrusted party, \emph{(iii)} all data
stored at and computation done at the collection agency,
\emph{(iv)} sharing data using data use agreements, and \emph{(v)} sharing
anonymized data.
%Question by JP: what do you mean by collection agency? Is it sequencing agency?

\subsubsection{Threat Model} We assume that data is being shared between
untrusted parties. The parties with whom data is being shared may want to use
it for any arbitrary purpose, including using it for participant re-identification, or for finding disease susceptibility of the patients or their blood relatives. 
%We also assume that once the data is shared, it can be used in an arbitrary manner.

\subsubsection{Solutions and Open Problems} We described some solutions in
Section~\ref{subsol:research}. However, these solutions do not allow for arbitrary
computations on encrypted data. Theoretically, many cryptographic solutions
exist to solve this issue. For example, fully homomorphic encryption (FHE) can be used
to encrypt the data and arbitrary computations can be done on it while
preserving privacy, but data needs to be decrypted by the party that encrypted
the data. Functional encryption (FE) could also be used, which allows
computation on encrypted data and produces plaintext directly. However, FHE and
FE are not sufficiently efficient to be practically useful. The performance of
FHE and FE is progressing and these schemes might be usable in the future to
support data sharing. Clearly though, specialized efficient solutions for
genomic data exploiting nature of genomic data are needed to support specific
analytics. Controlled Functional Encryption (C-FE)~\cite{Naveed2014CFE}
described above is a promising approach to develop practical solutions.

\ignore{
\subsection{Genome Annotation}
\TODO{Needs help with this section}

\subsubsection{Setting}

\subsubsection{Threat Model}

\subsubsection{Solutions and Open Problems}
}

\subsection{Paternity}
Genomic data is extensively used to determine parentage and test results are admissible
in courts of law. Today, the biospecimen of the individuals involved in the tests are outsourced to
a third party in the form of cheek swabs, where the DNA is extracted.
%conduct the test.
Sending one's DNA to a third party could have serious implications for one's privacy.
%If an individual has their genome sequenced beforehand, they would be able to compute paternity in privacy-preserving manner.

\subsubsection{Setting}
Two parties each have their sequenced genome or genotyped variants and one party wants to know whether the other party is the parent.

\subsubsection{Threat Model}
The threat model in this case is the standard model defined for secure
two-party computations. We assume that parties can be honest-but-curious or
malicious.

\subsubsection{Solutions and Open Problems}
In Section~\ref{sec:solutions}, we explain some of the solutions to the problem. A
chemical test -- RFLP -- can be simulated for a neat and efficient
privacy-preserving solution, given that genomes are stored by individuals
themselves~\cite{Baldi2011}. Yao's garbled circuits can be used instead of PSI
to output a binary answer (YES or NO) instead of the number of matched segments in
simulated RFLP test and therefore reveals the minimum amount of information.
%Due to privacy concerns explained in Section~\ref{sec:riskassess},
%genomes might be stored at a hospital or a third party.

\subsection{Forensic DNA Databases}
Many countries maintain a huge database of DNA profiles of convicted (and, in some cases, accused)
criminals. Law enforcement agencies usually have unlimited access to such a resource,
which makes it vulnerable to abuse. It is possible that in the near future, instead
of concise DNA profiles, law enforcement agencies will be able to have access
to full genome sequences of individuals, which further exacerbates the issues.

\subsubsection{Setting}
Police officers collect DNA samples from a crime scene. Then, they
want to check whether an individual with the same DNA profile/sequence is present
in the DNA records database.

\subsubsection{Threat Model}
We assume that the adversary can be honest-but-curious, interested in learning
about other people in the database. In addition, the adversary can be
malicious. If the adversary has write access to the database, he can also try
to compromise the integrity of the record(s) in the database. We also assume
that the adversary is able to affect the outcome of a query in an arbitrary
manner.

\subsubsection{Solutions and Open Problems}
We discussed some of the existing solutions to this problem in Section~\ref{sec:solutions}. Theoretically, this problem differs from
interpretation and other types of computation, as the privacy for query is not
required, only the privacy of the individuals other than the suspect is of
concern here. This makes the problem more tractable, possibly making
solutions scalable to large databases with millions of records.

\subsection{Recreational Genomics}
Several commercial companies offer direct-to-consumer genomics services. They
include kinship, ancestry, and partner compatibility testing.

\subsubsection{Setting}
The customer ships her specimen (usually saliva) to a company. This specimen is
used to genotype typically one million SNPs
%using microarray technology
and the
data is then digitized and stored in digital form on the server. Some
computation is done on this data for ancestry, disease susceptibility, kinship, or other tests.  The data and the results are then available for the user to
download or see through a browser. Some companies (e.g., 23andme) allow
to search for people with common ancestors e.g., 3rd, 4th or 5th cousins.
%Some
%customers put this data online (e.g., on genome-sharing websites like OpenSNP.org) and some
%even publish their name, picture and detailed phenotypic information along
%with their genomic data.

\subsubsection{Threat Model}
We assume the threat models for specimen collection, digitization and
interpretation.
%are the same as discussed
%earlier.
%Section~\ref{ssec:biospecimen} and Section~\ref{ssec:digitization}
%respectively, so same threat models are valid here. The types of computation done
%in recreational genomics are also similar to previous section.
%\vspace{-5pt}
There are also new threats. The owner of the data posts his data online along
with his identity and some phenotypical data, as done for example on
openSNP~\cite{greshake2014opensnp}. We assume that the data owner makes an informed decision, so he
willingly gives away his genome privacy. The major threat then is to the blood
relatives of the data owner, whose private information is also leaked.
%(as
%discussed in Section~\ref{sec:risk_DTC}).

\indent \textit{Genomic data aggregation} is another issue caused by users posting their
genomic data online. This data can be aggregated by companies and then used
for commercial or other purposes. It is worth noting that some genome-sharing
websites have
achievement programs where users get rewards whenever they post their
phenotype data (e.g., hair color, eye color, disease pre-disposition, etc.).
Genomic data along with phenotypic data is much more valuable than genomic
data alone.

\subsubsection{Solutions and Open Problems}
All other solutions apply here, however recreational genomics
presents new problems of public sharing and genomic data aggregation. We emphasize that public
awareness is required to enable people to make an informed decision to share
their genomic data publicly because such sharing compromises their own and their relatives' privacy.
It should be made clear that in case of abuse of this publicly available data,
people would be discouraged to share genomic data even for legitimate research
purposes. However, this is a policy and ethics debate and is out
of scope of this paper.

\ignore{
\subsection{Research}

\subsection{Healthcare}

\subsection{Legal and Forensic}

\subsection{Direct-to-consumer}
}

\ignore{
\subsection{Datasets}
\label{ssec:udatasets}

\TODO{Designing a self-contained and self-explanatory API for querying public
	or private genomic databases for privacy enhancing technologies such that it
		requires minimal biology background.}
%\TODO{Understanding genomics datasets made easy}
}
﻿\section*{Conclusion}
%The collection and processing of genomic data is exploding at an exponential
%rate.
The confluence of cheap computing and high-throughput sequencing technologies
is making genomic data increasingly easy to collect, store, and process.
%As this information becomes more widely available for computational purposes
%Given its distinguishing features (e.g., influence on health and kinship
%relations), and its application in a wide range of setting is have detailed
%the essential features of genomic data and explained its multiple uses.
%Leveraging on these facts, we have discussed the relevance of genomic
At the same time, genomic data is being integrated into a wide range of
applications in diverse settings (e.g., healthcare, research, forensics, direct-to-consumer), such that privacy and security issues
%associated with genomic data
have yet to be sufficiently defined and addressed.
%and provided
%the outcome of an expert survey.
%More and more, genomics (and healthcare in general) is relying on computers.
For instance, massive computation capability is needed
%on terabytes
to analyze genomic data for research purposes, such that the cloud is likely to play a large role in the management of such data.
%of data to be able to associate a given set of genes to a health condition.
At the same time, genomic data will be used in specific applications (e.g.,
		forensics) where mobile computing environments (e.g., tablets and
			smartphones) will be routinely used to access those data. And, genomic
		data will be
increasingly available on the Web, especially in citizen-contributed environments, (e.g., online social networks).
%Patients (and people in general) will expect an increased availability of this data, without jeopardizing their privacy.
While some individuals are sharing such information,
%Much more is on the way: substantial progress will happen
there are significant privacy concerns because it is unknown what such
information %will
is capable of revealing, or how it will be used in the future.\\
%in the coming years in the areas of metabolomics, proteomics, and transcriptomics, paving the way to unprecedented breakthroughs in medicine, but also (unavoidably) revealing very intimate data about everyone's body.
\indent As such, there is a clear need to support personalized medicine, genomic research, forensic investigation, and recreational genomics while respecting privacy. Computing is a crucial enabler, but can also be the first
source of leakage if appropriate mechanisms are not put in place.
%We hope this systematization of knowledge will help the CS community to more assertively address this challenge.
Our survey (opinion poll from the biomedical community)
%illustrates what users of such data (particularly biomedical researchers)
	provides some insight into what may be the most important aspects of the problem to study.  And, along these lines, we have provided
a review of the state-of-the-art regarding computational protection methods in this
%(still embryonic)
field, as well as the main challenges moving forward. To assist the data privacy and security community to develop meaningful solutions, we have provided a framework to facilitate the understanding of the privacy-preserving
handling of genomic data.

\bibliographystyle{natbib}
\bibliography{main}

\appendix
%\appendixhead{NAVEED}
\section*{APPENDIX}
In this appendix, we provide additional results from our expert opinion
poll discussed in Section~\ref{sec:survey}. Results shown here are stratified
according to the security/privacy and genetics/genomics expertise of the
participants. Results for the cases where the number of participants was small
are omitted.

\begin{figure}[h]
 \centering
 \includegraphics[width=1\textwidth]{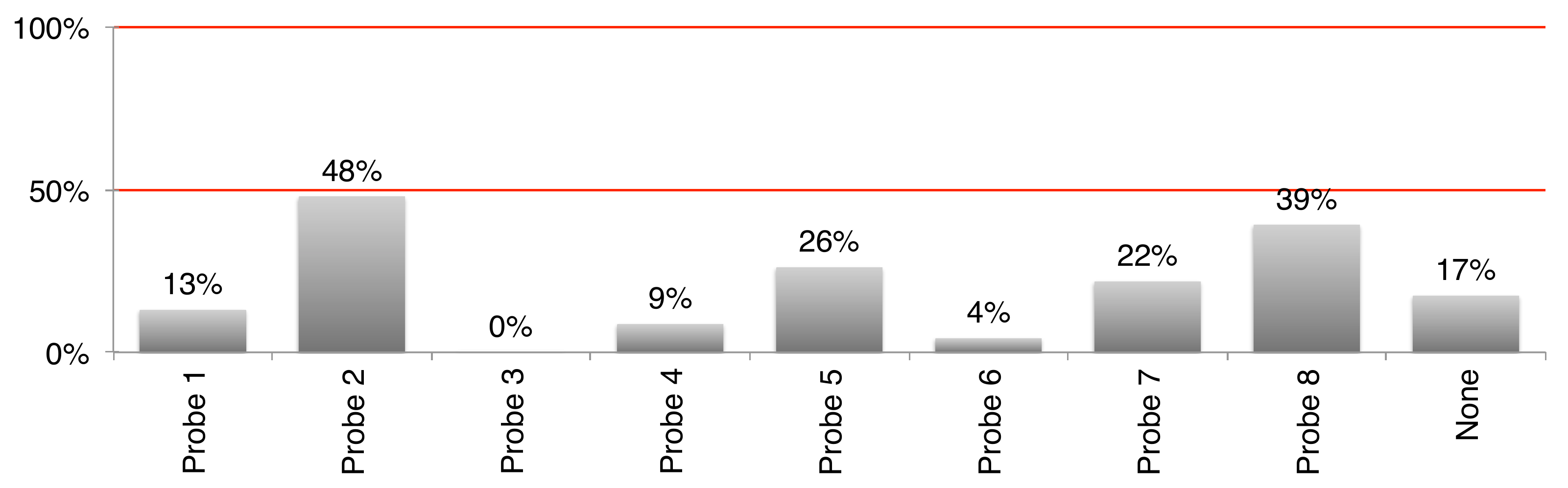}
 \caption{Response to the question: \emph{Do you believe that: (Multiple options can be checked)}. The probes are described in detail in Figure~\ref{fig:beliefstext}. ``None'' means the respondent does not agree with any of the probes:
 \textbf{Only ``Expert'' biomedical participants (sample size 23).}}
 \label{fig:beliefs_expert}
 \vspace{-0.5cm}
\end{figure}
\begin{figure}[h]
 \centering
 \includegraphics[width=1\textwidth]{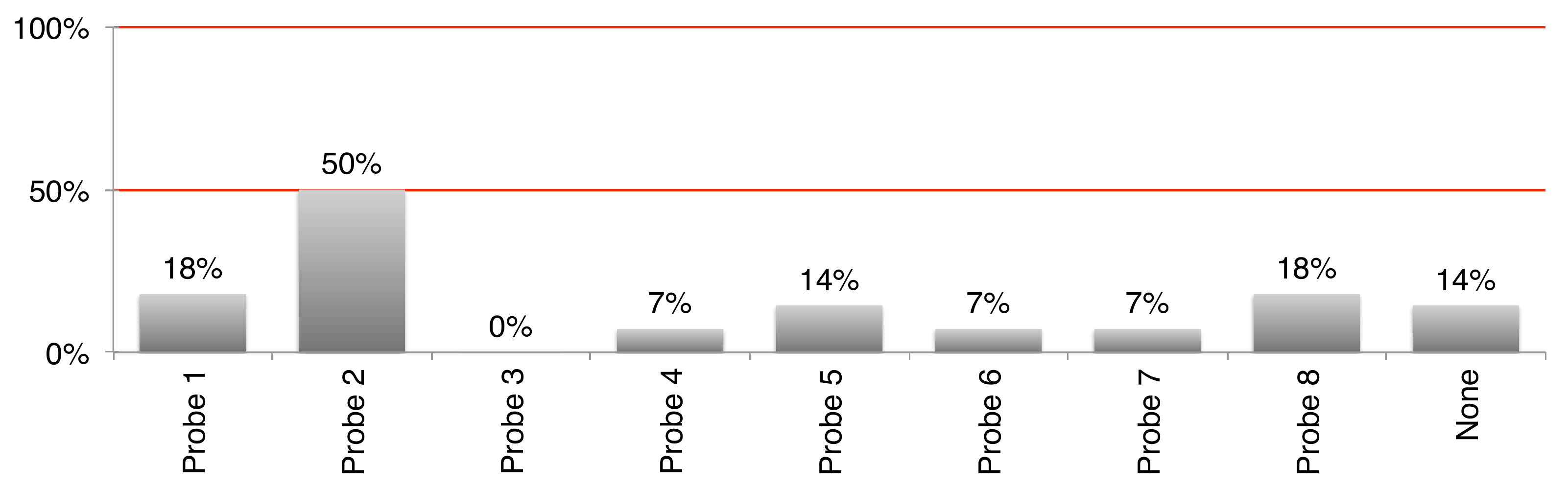}
 \caption{Response to the question: \emph{Do you believe that: (Multiple options can be checked)}. The probes are described in detail in Figure~\ref{fig:beliefstext}. ``None'' means the respondent does not agree with any of the probes:
 \textbf{Only ``Knowledgeable'' biomedical participants (sample size 28).}}
 \label{fig:beliefs_know}
 \vspace{-0.5cm}
\end{figure}
\begin{figure}[h]
 \centering
 \includegraphics[width=1\textwidth]{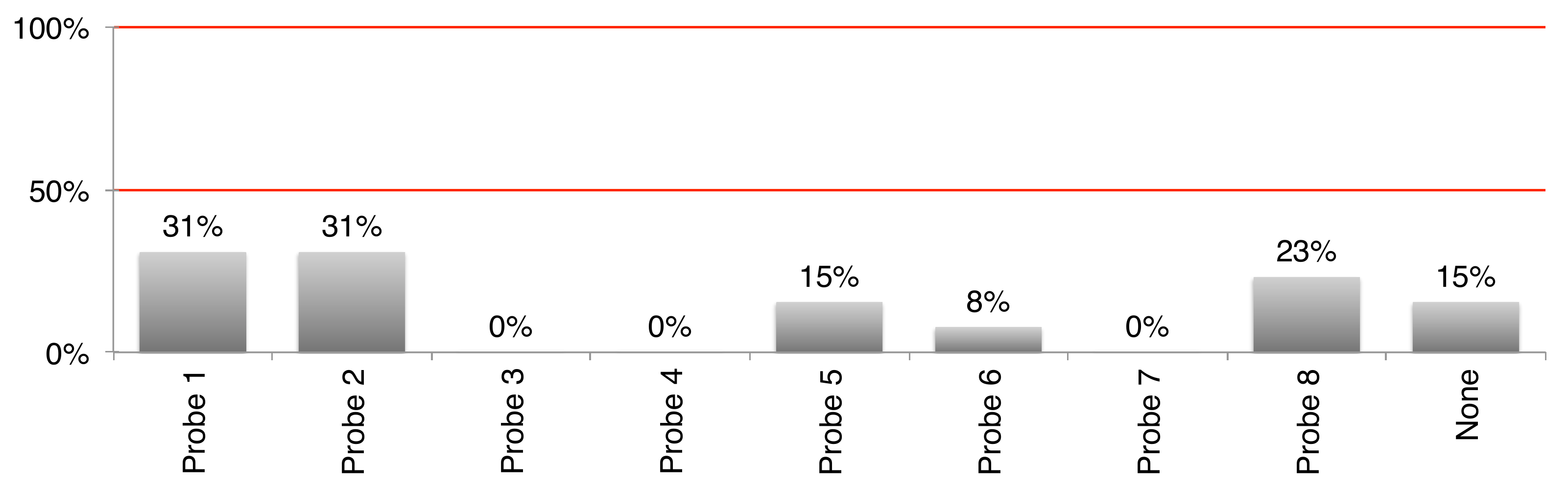}
 \caption{Response to the question: \emph{Do you believe that: (Multiple options can be checked)}. The probes are described in detail in Figure~\ref{fig:beliefstext}. ``None'' means the respondent does not agree with any of the probes:
 \textbf{Only ``Some familiarity'' biomedical participants (sample size 13).}}
 \label{fig:beliefs_some}
 \vspace{-0.5cm}
\end{figure}
\begin{figure}[h]
 \centering
 \includegraphics[width=1\textwidth]{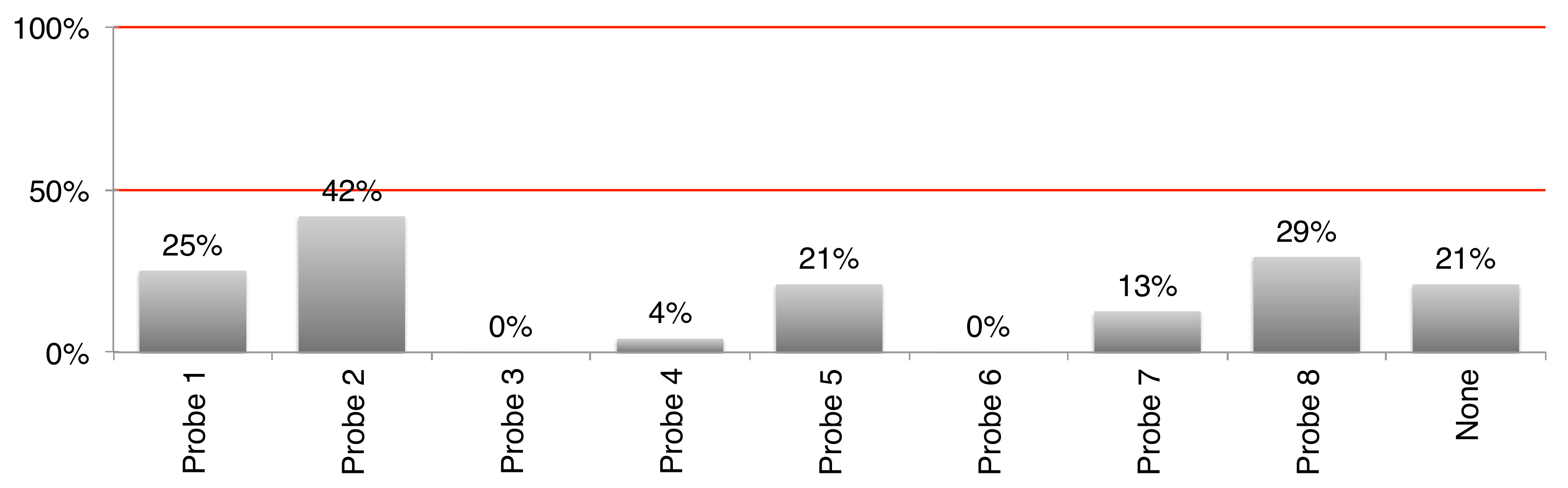}
 \caption{Response to the question: \emph{Do you believe that: (Multiple options can be checked)}. The probes are described in detail in Figure~\ref{fig:beliefstext}. ``None'' means the respondent does not agree with any of the probes:
 \textbf{Only ``Knowledgeable'' security and privacy participants (sample size 24).}}
 \label{fig:beliefs_know_sp}
 \vspace{-0.5cm}
\end{figure}
\begin{figure}[h]
 \centering
 \includegraphics[width=1\textwidth]{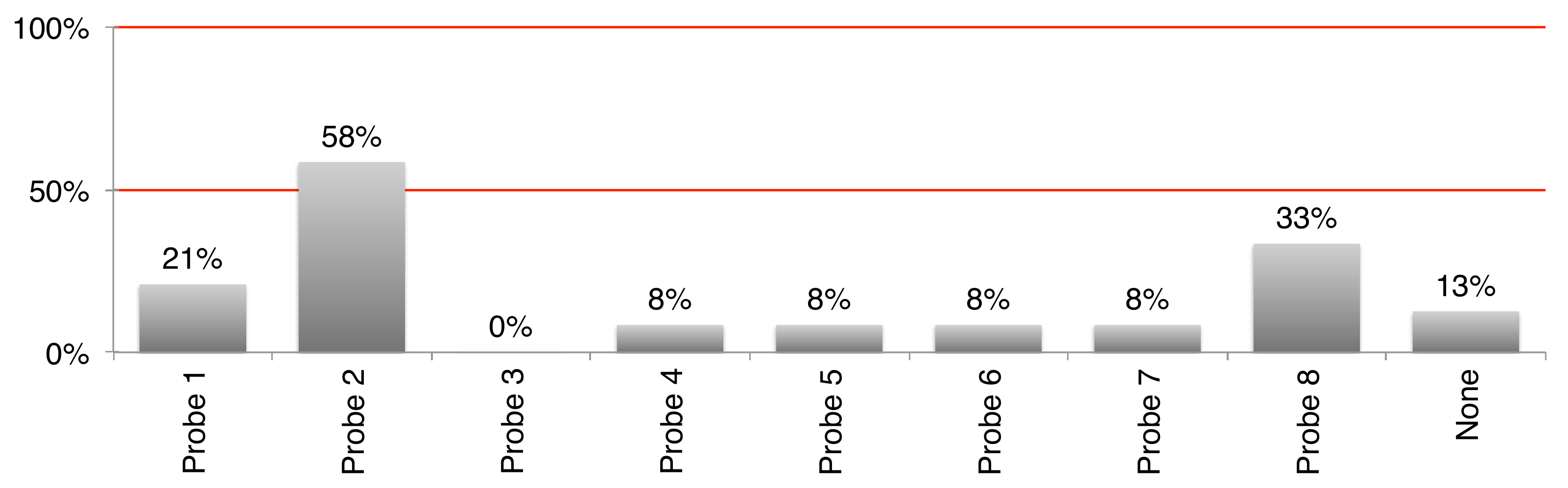}
 \caption{Response to the question: \emph{Do you believe that: (Multiple options can be checked)}. The probes are described in detail in Figure~\ref{fig:beliefstext}. ``None'' means the respondent does not agree with any of the probes:
\textbf{Only ``Some familiarity'' security and privacy participants (sample size 24).}}
 \label{fig:beliefs_some_sp}
 \vspace{-0.5cm}
\end{figure}
\begin{figure}[h]
 \centering
 \includegraphics[width=1\textwidth]{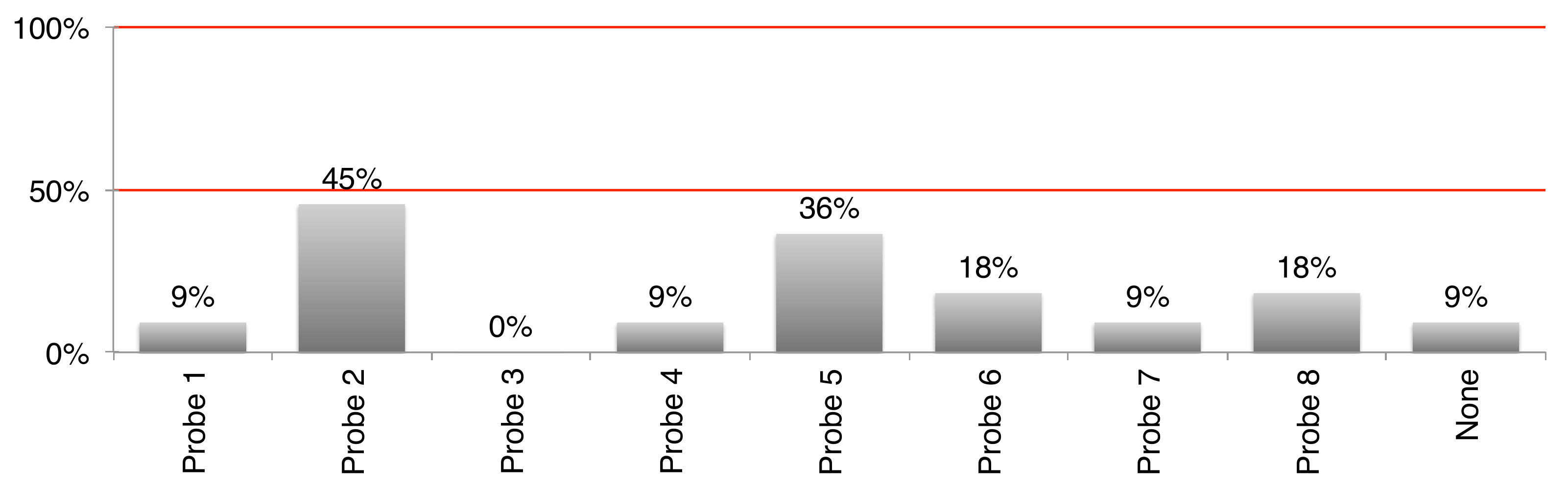}
 \caption{Response to the question: \emph{Do you believe that: (Multiple options can be checked)}. The probes are described in detail in Figure~\ref{fig:beliefstext}. ``None'' means the respondent does not agree with any of the probes:
\textbf{Only ``No familiarity'' security and privacy participants (sample size 11).}}
 \label{fig:beliefs_no_sp}
 \vspace{-0.5cm}
\end{figure}
\begin{figure}[h]
 \centering
 \includegraphics[width=1\textwidth]{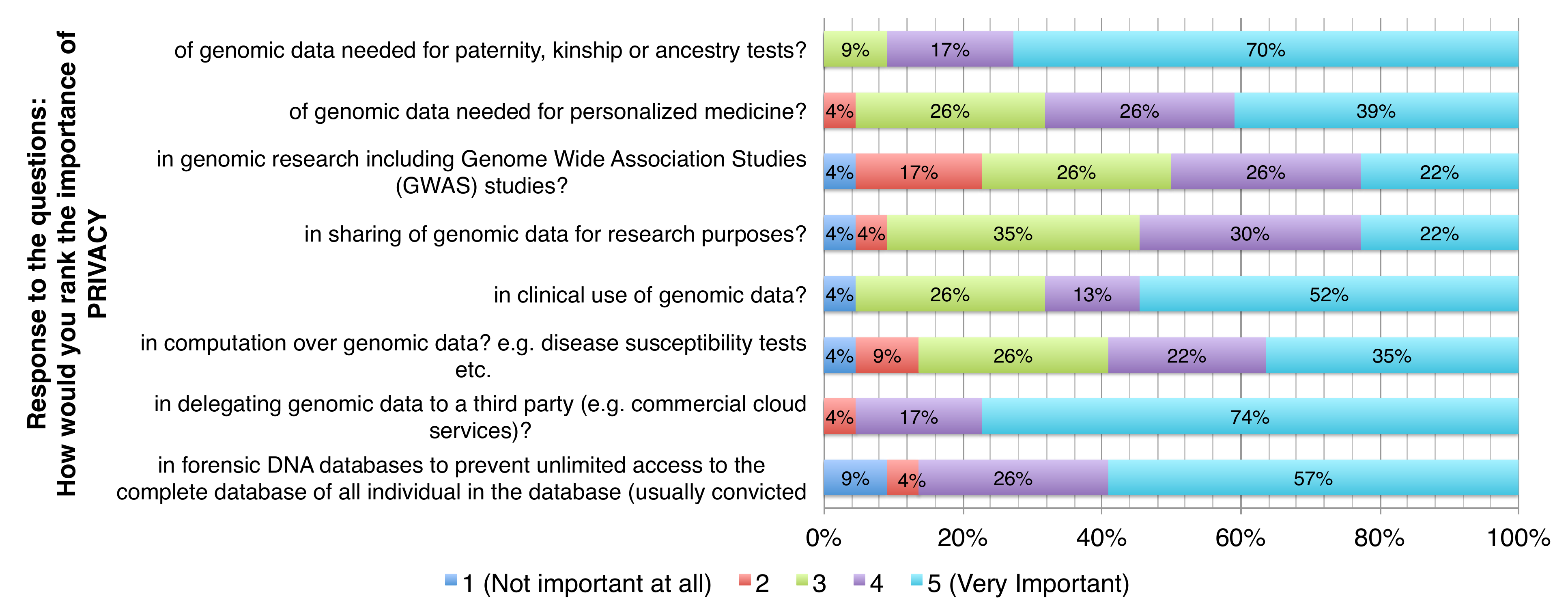}
 \caption{Relevance of genome privacy research done by the computer science
 community: \textbf{Only ``Expert'' biomedical participants (sample size 23).}}
 \label{fig:privacyquestions_expert}
 \vspace{-1cm}
\end{figure}
\begin{figure}[h]
 \centering
 \includegraphics[width=1\textwidth]{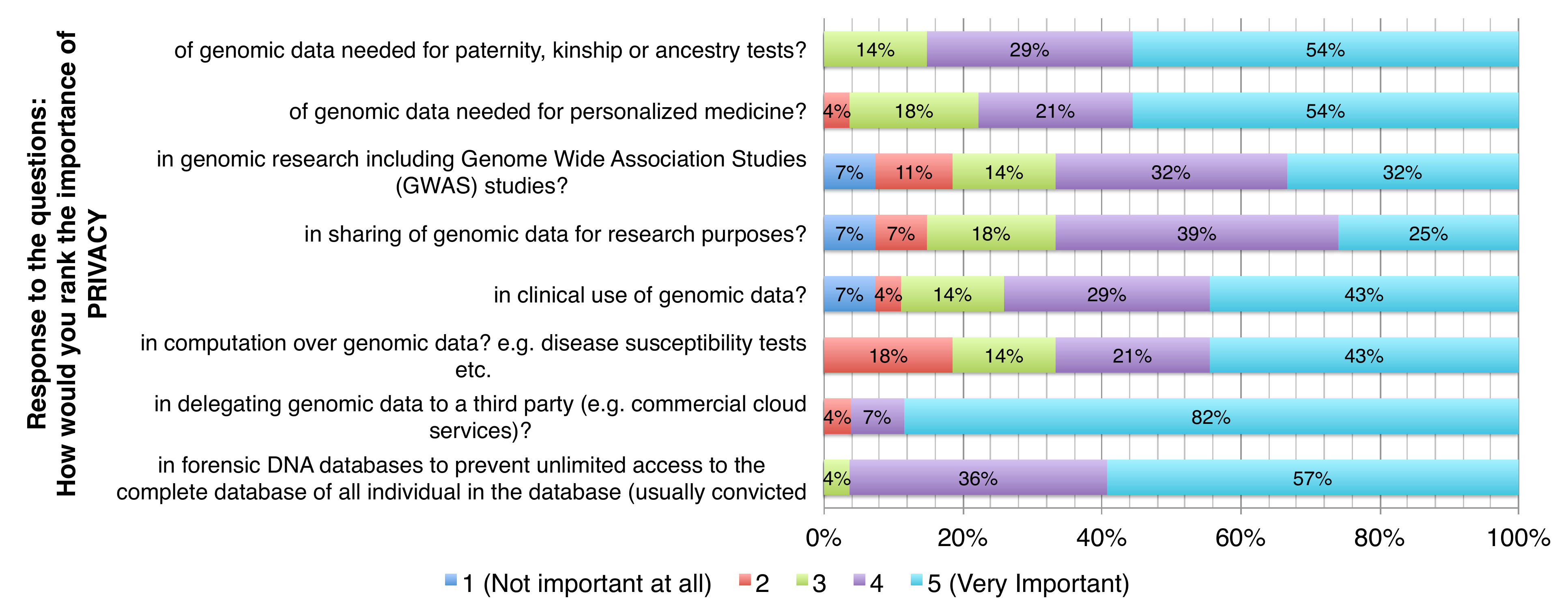}
 \caption{Relevance of genome privacy research done by the computer science
 community: \textbf{Only ``Knowledgeable'' biomedical participants (sample size 28).}}
 \label{fig:privacyquestions_know}
 \vspace{-0.5cm}
\end{figure}
\begin{figure}[h]
 \centering
 \includegraphics[width=1\textwidth]{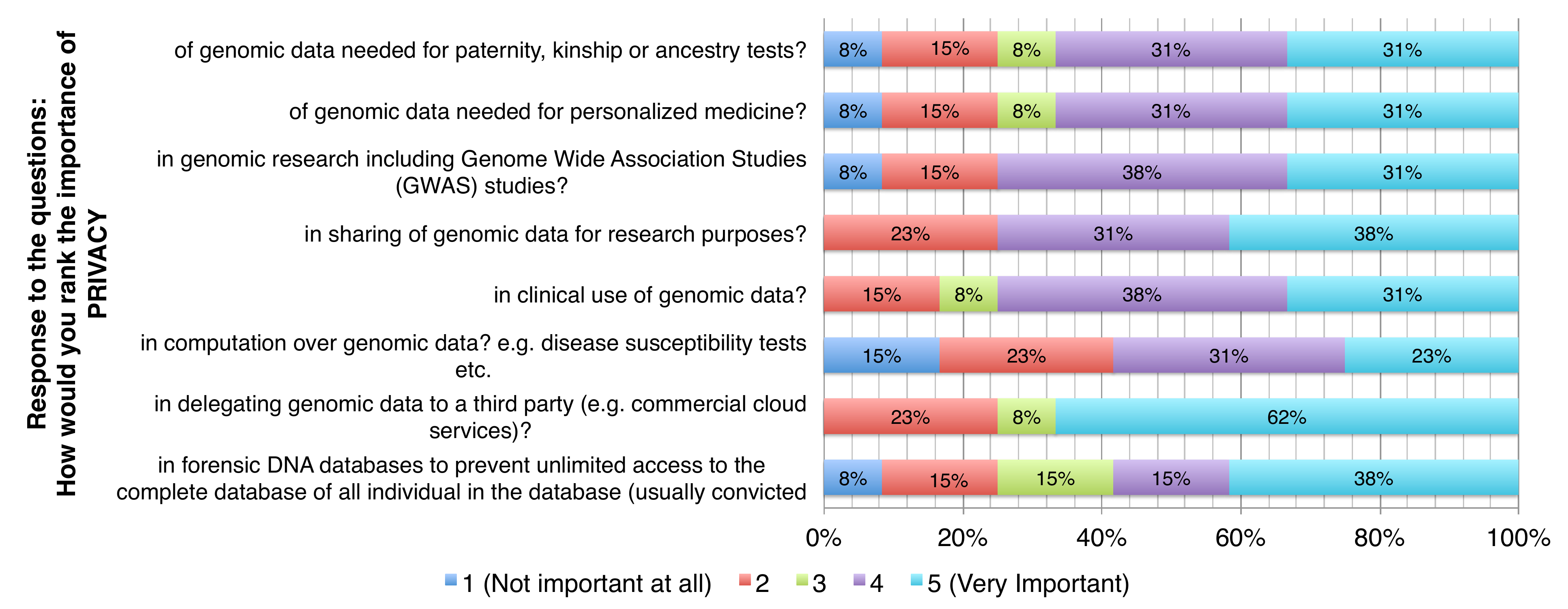}
 \caption{Relevance of genome privacy research done by the computer science
 community: \textbf{Only ``Some familiarity'' biomedical participants (sample size 13).}}
 \label{fig:privacyquestions_some}
 \vspace{-0.5cm}
\end{figure}
\begin{figure}[h]
 \centering
 \includegraphics[width=1\textwidth]{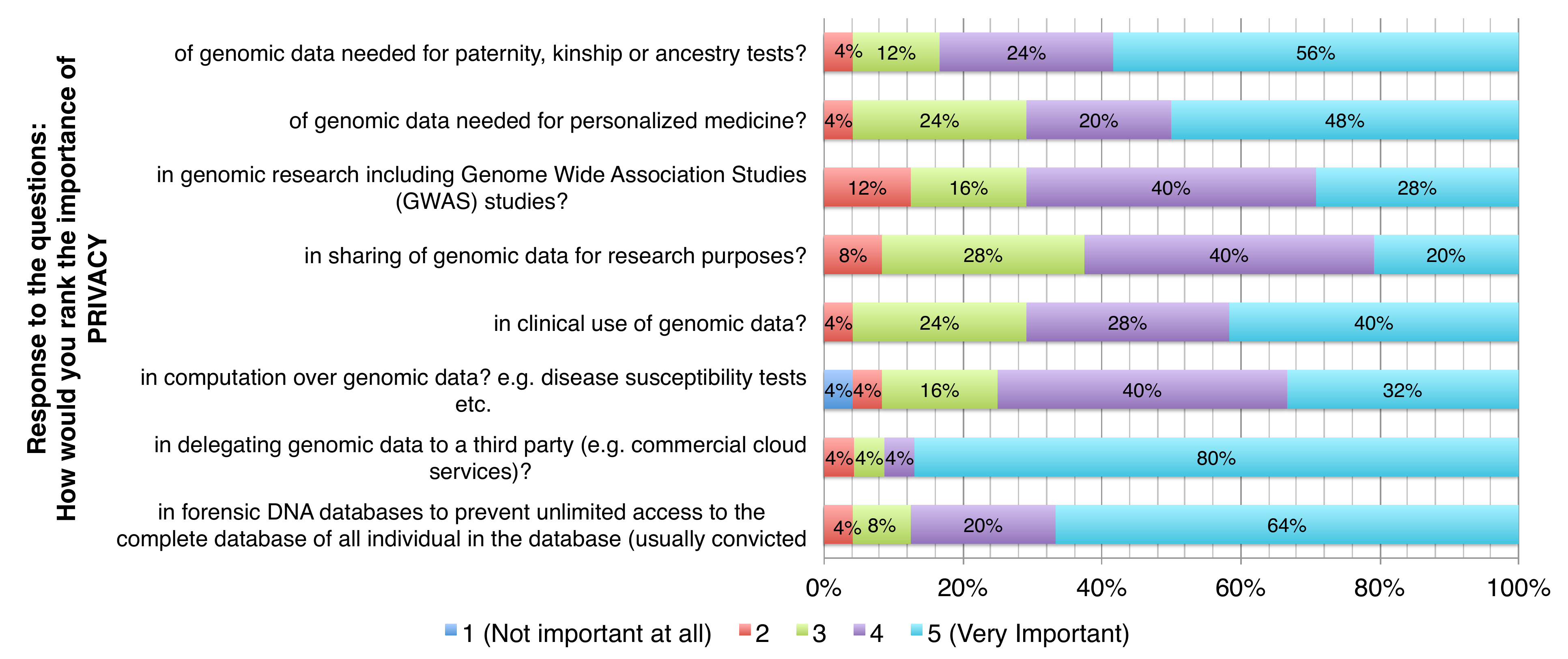}
 \caption{Relevance of genome privacy research done by the computer science
 community: \textbf{Only ``Knowledgeable'' security and privacy participants (sample size 24).}}
 \label{fig:privacyquestions_know_sp}
 \vspace{-0.5cm}
\end{figure}
\begin{figure}[h]
 \centering
 \includegraphics[width=1\textwidth]{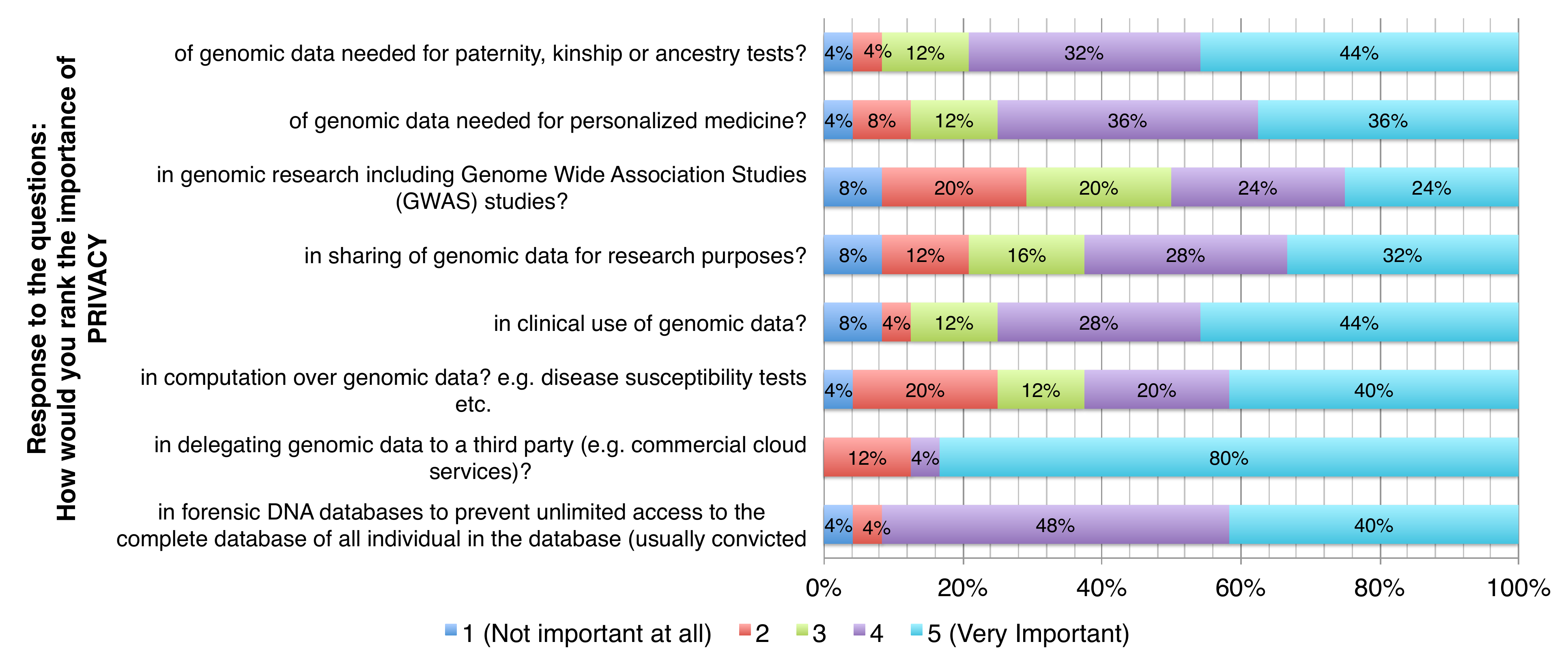}
 \caption{Relevance of genome privacy research done by the computer science
 community: \textbf{Only ``Some familiarity'' security and privacy participants (sample size 24)}.}
 \label{fig:privacyquestions_some_sp}
 \vspace{-0.5cm}
\end{figure}
\begin{figure}[h]
 \centering
 \includegraphics[width=1\textwidth]{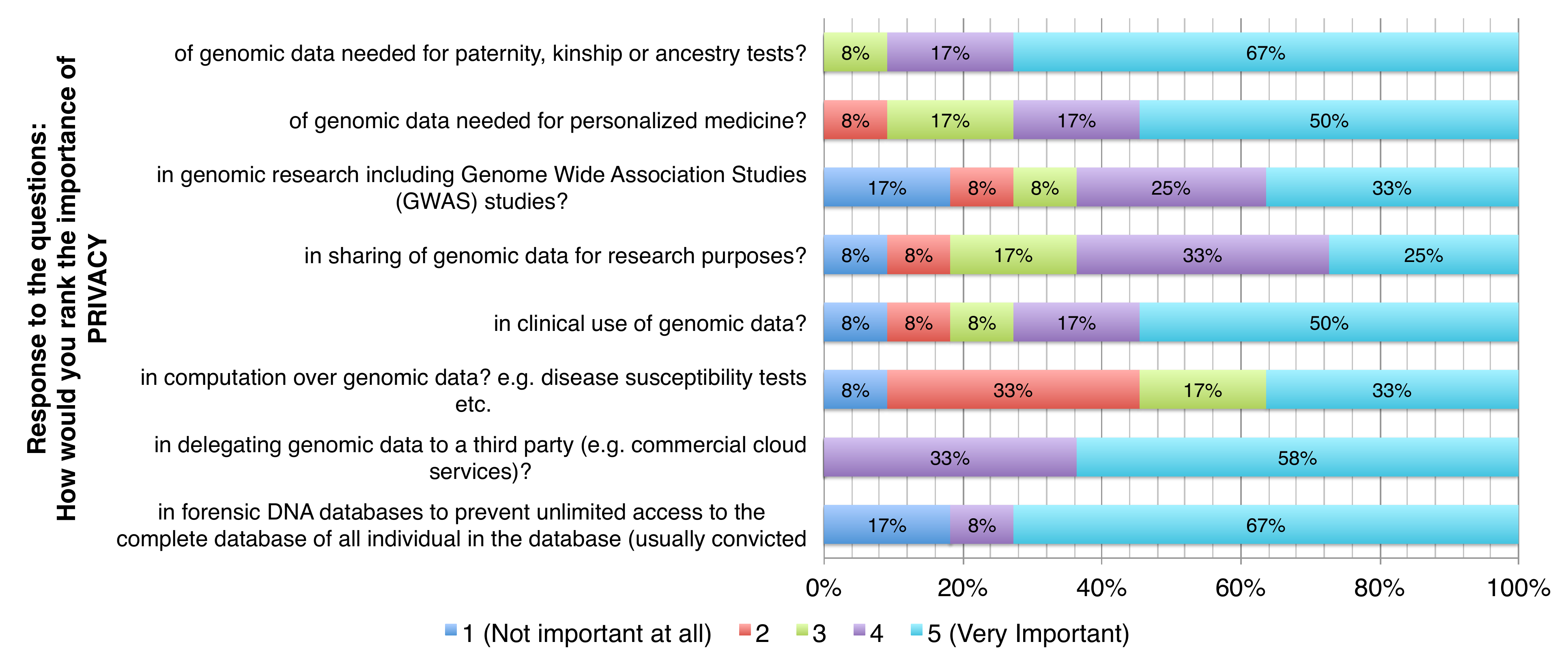}
 \caption{Relevance of genome privacy research done by the computer science
 community: \textbf{Only ``No familiarity'' security and privacy participants (sample size 11).}}
 \label{fig:privacyquestions_no_sp}
 \vspace{-0.5cm}
\end{figure}
\begin{figure}[h]
 \centering
 \includegraphics[width=1\textwidth]{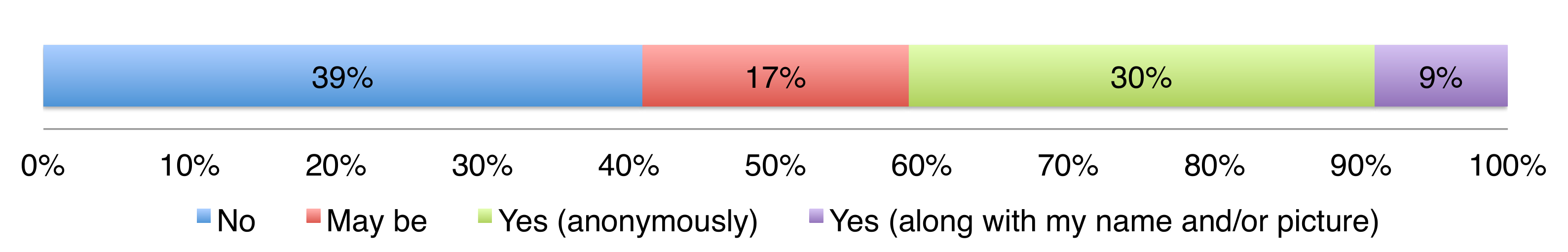}
\caption{Response to the question: \emph{Would you publicly share your genome on the Web?}: 
\textbf{Only ``Expert'' biomedical participants (sample size 23).}}
 \label{fig:kinship_expert}
 \vspace{-0.5cm}
\end{figure}
\begin{figure}[h]
 \centering
 \includegraphics[width=1\textwidth]{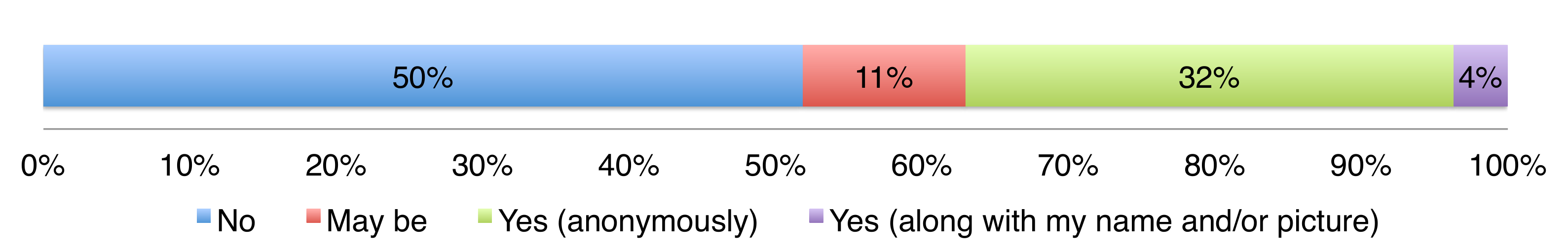}
\caption{Response to the question: \emph{Would you publicly share your genome on the Web?}: 
 \textbf{Only ``Knowledgeable'' biomedical participants (sample size 28).}}
 \label{fig:kinship_know}
 \vspace{-0.5cm}
\end{figure}
\begin{figure}[h]
 \centering
 \includegraphics[width=1\textwidth]{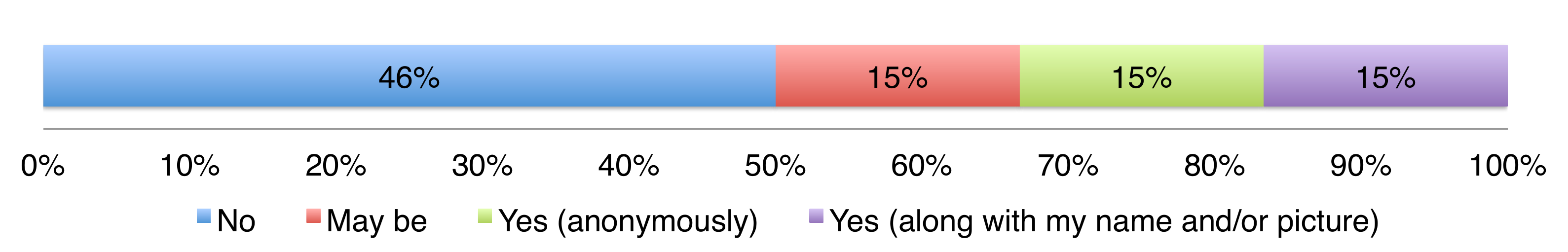}
\caption{Response to the question: \emph{Would you publicly share your genome on the Web?}: 
 \textbf{Only ``Some familiarity'' biomedical participants (sample size 13).}}
 \label{fig:kinship_some}
 \vspace{-0.5cm}
\end{figure}
\begin{figure}[h]
 \centering
 \includegraphics[width=1\textwidth]{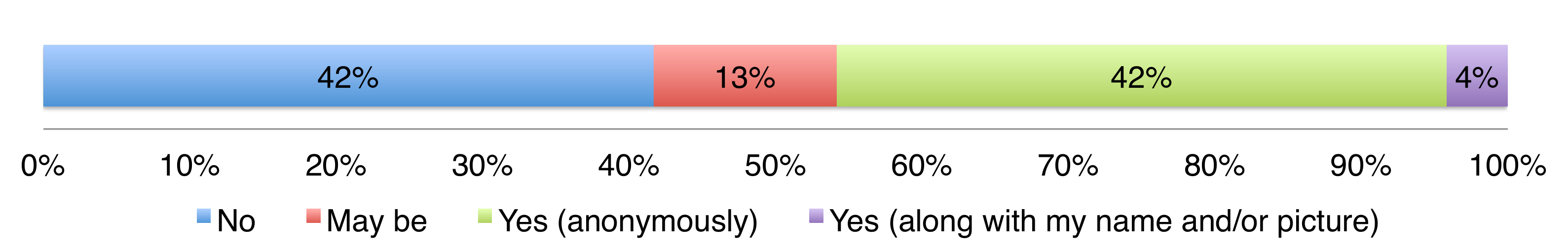}
\caption{Response to the question: \emph{Would you publicly share your genome on the Web?}: 
 \textbf{Only ``Knowledgeable'' security and privacy participants (sample size 24).}}
 \label{fig:kinship_know_sp}
 \vspace{-0.5cm}
\end{figure}
\begin{figure}[h]
 \centering
 \includegraphics[width=1\textwidth]{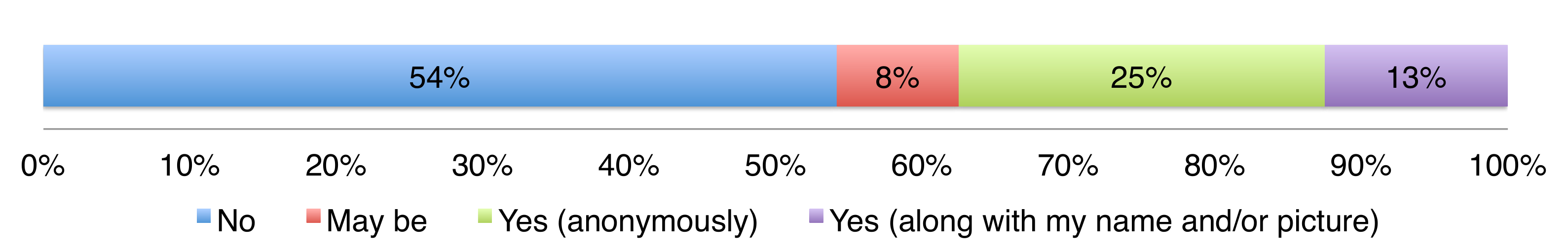}
\caption{Response to the question: \emph{Would you publicly share your genome on the Web?}: 
 \textbf{Only ``Some familiarity'' security and privacy participants (sample size 24).}}
 \label{fig:kinship_some_sp}
 \vspace{-0.5cm}
\end{figure}
\begin{figure}[h]
 \centering
 \includegraphics[width=1\textwidth]{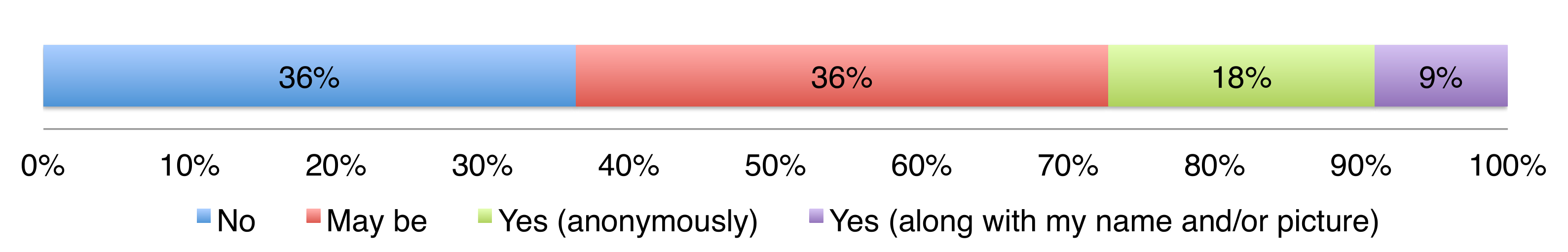}
\caption{Response to the question: \emph{Would you publicly share your genome on the Web?}: 
 \textbf{Only ``No familiarity'' security and privacy participants (sample size 11).}}
 \label{fig:kinship_no_sp}
 \vspace{-0.5cm}
\end{figure}
\begin{figure}[h]
 \centering
 \includegraphics[width=1\textwidth]{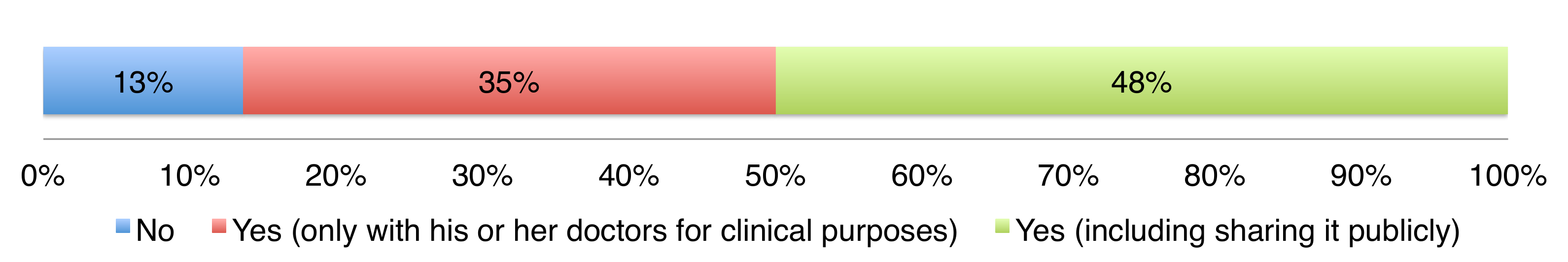}
\caption{Response to the question: \emph{Assuming that one's genomic data leaks a lot of private information about his or her relatives, do you think one should have the right to share his or her genomic data?}:
\textbf{Only ``Expert'' biomedical participants (sample size 23).}}
 \label{fig:sharing_expert}
 \vspace{-0.5cm}
\end{figure}
\begin{figure}[h]
 \centering
 \includegraphics[width=1\textwidth]{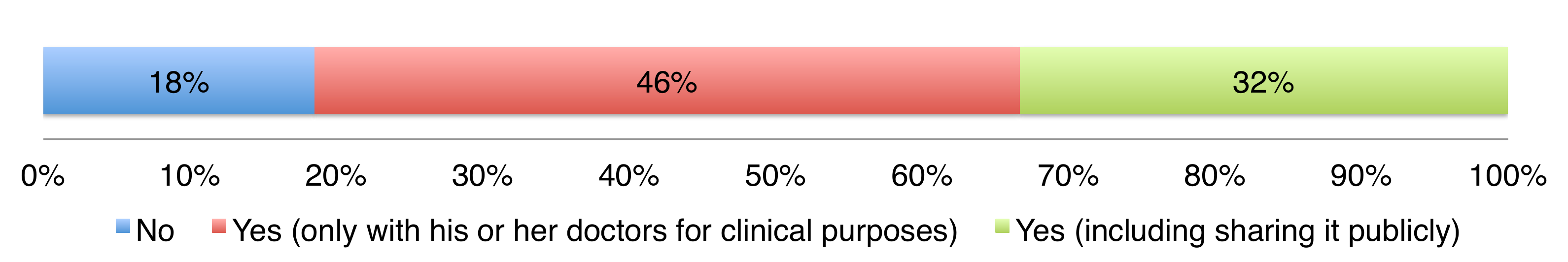}
\caption{Response to the question: \emph{Assuming that one's genomic data leaks a lot of private information about his or her relatives, do you think one should have the right to share his or her genomic data?}:
 \textbf{Only ``Knowledgeable'' biomedical participants (sample size 28).}}
 \label{fig:sharing_know}
 \vspace{-0.5cm}
\end{figure}
\begin{figure}[h]
 \centering
 \includegraphics[width=1\textwidth]{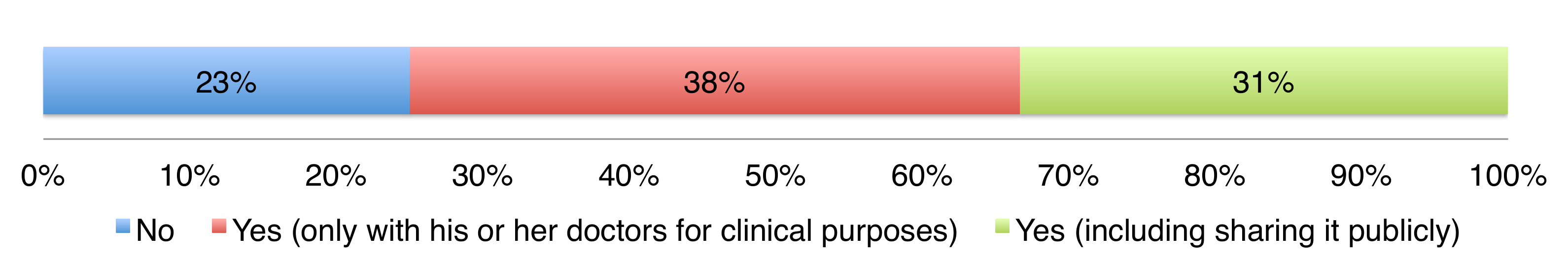}
\caption{Response to the question: \emph{Assuming that one's genomic data leaks a lot of private information about his or her relatives, do you think one should have the right to share his or her genomic data?}:
 \textbf{Only ``Some familiarity'' biomedical participants (sample size 13).}}
 \label{fig:sharing_some}
 \vspace{-0.5cm}
\end{figure}
\begin{figure}[h]
 \centering
 \includegraphics[width=1\textwidth]{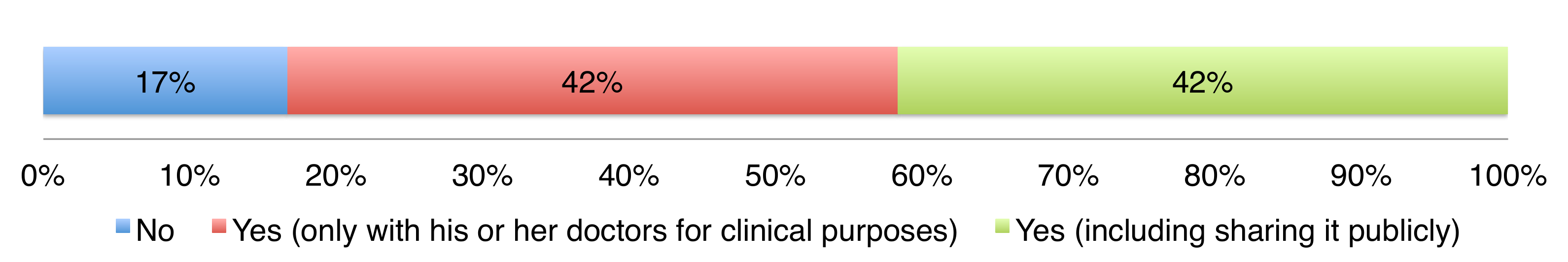}
\caption{Response to the question: \emph{Assuming that one's genomic data leaks a lot of private information about his or her relatives, do you think one should have the right to share his or her genomic data?}:
 \textbf{Only ``Knowledgeable'' security and privacy participants (sample size 24).}}
 \label{fig:sharing_know_sp}
 \vspace{-0.5cm}
\end{figure}
\begin{figure}[h]
 \centering
 \includegraphics[width=1\textwidth]{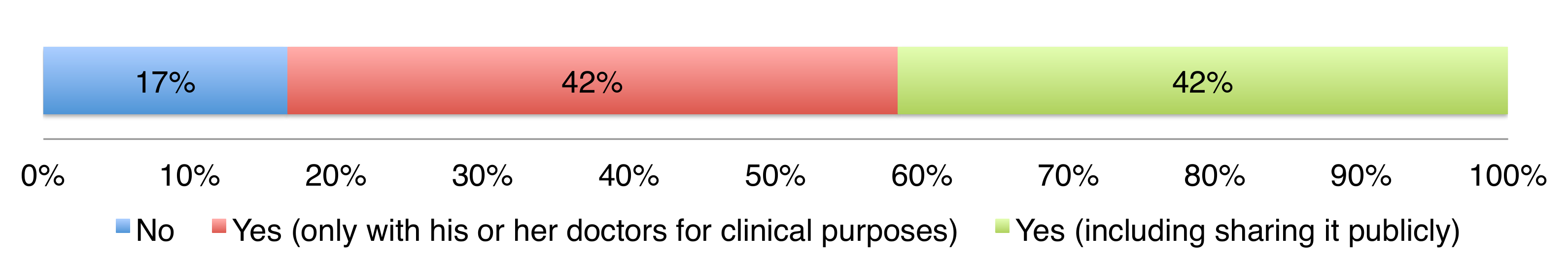}
\caption{Response to the question: \emph{Assuming that one's genomic data leaks a lot of private information about his or her relatives, do you think one should have the right to share his or her genomic data?}:
 \textbf{Only ``Some familiarity'' security and privacy participants (sample size 24).}}
 \label{fig:sharing_some_sp}
 \vspace{-0.5cm}
\end{figure}
\begin{figure}[h]
 \centering
 \includegraphics[width=1\textwidth]{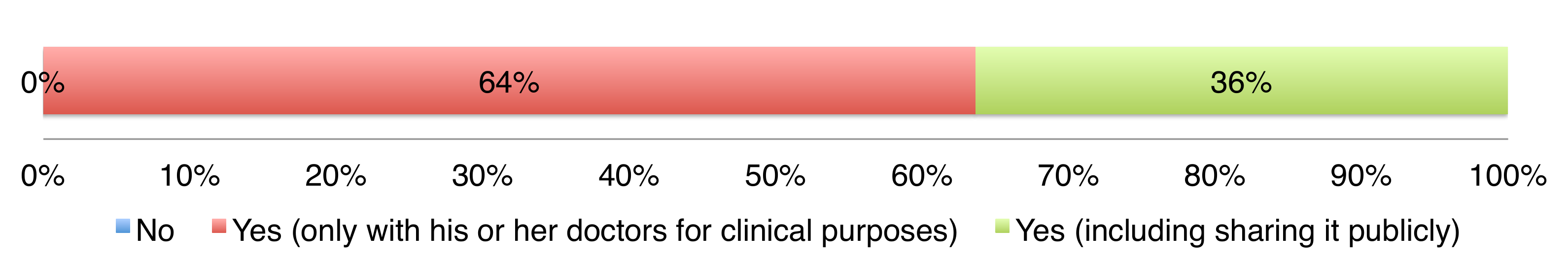}
\caption{Response to the question: \emph{Assuming that one's genomic data leaks a lot of private information about his or her relatives, do you think one should have the right to share his or her genomic data?}:
 \textbf{Only ``No familiarity'' security and privacy participants (sample size 11).}}
 \label{fig:sharing_no_sp}
 \vspace{-0.5cm}
\end{figure}

\end{document}